\documentclass[sigconf]{acmart}
\usepackage{subcaption}
\usepackage{adjustbox}
\usepackage{multirow}
\AtBeginDocument{%
  }

\setcopyright{acmlicensed}

\copyrightyear{2026}
\acmYear{2026}
\setcopyright{cc}
\setcctype{by}
\acmConference[MM '26] {Proceedings of the 34th ACM International Conference on Multimedia}{November 10--14, 2026}{Rio de Janeiro, Brazil.}
\acmBooktitle{Proceedings of the 34th ACM International Conference on Multimedia (MM '26), November 10--14, 2026, Rio de Janeiro, Brazil}
\acmISBN{979-8-4007-2213-4/2026/11}

\acmDOI{ 10.1145/3767308.3835957}

\acmISBN{979-8-4007-2213-4/2026/11}

\begin{document}

\title{Multimodal Decoupling for Diversified Recommendation through Semantic and Structural Shaping}

\author{Changran He}

\affiliation{%
  \institution{Harbin Institute of Technology(Shenzhen)}
  \city{Shenzhen}
  \country{China}
}
\email{2023311c12@stu.hit.edu.cn}
\authornote{Both authors contributed equally to this research.}


\author{Jiaming Wei}

\affiliation{%
  \institution{Harbin Institute of Technology(Shenzhen)}
  \city{Shenzhen}
  \country{China}
}
\email{23s151033@stu.hit.edu.cn}
\authornotemark[1]

\author{Jianghong Ma}
\authornote{Corresponding author.}

\affiliation{%
  \institution{Shenzhen Key Laboratory of Internet Information Collaboration}
  \city{Shenzhen}
  \country{China}
}

\affiliation{%
  \institution{\mbox{Harbin Institute of Technology(Shenzhen)}}
  \city{Shenzhen}
  \country{China}
}

\affiliation{%
  \institution{Shenzhen Research Institute,
  City University of Hong Kong}
  \city{Shenzhen}
  \country{China}
}
\email{majianghong@hit.edu.cn}
\email{jianghma2-c@my.cityu.edu.hk}

\author{Tianjun Wei}

\affiliation{%
  \institution{Nanyang Technological University}
  \city{Singapore}
  \country{Singapore}
}
\email{tjwei2-c@my.cityu.edu.hk}

\author{Haijun Zhang}
\affiliation{%
  \institution{Harbin Institute of Technology(Shenzhen)}
  \city{Shenzhen}
  \country{China}
}
\email{hjzhang@hit.edu.cn}

\author{Xiaofeng Zhang}
\affiliation{%
  \institution{Harbin Institute of Technology(Shenzhen)}
  \city{Shenzhen}
  \country{China}
}
\email{zhangxiaofeng@hit.edu.cn}

\renewcommand{\shortauthors}{Changran He et al.}

\begin{abstract}
  Recommender systems often suffer from over-concentrated results when optimized for accuracy, limiting recommendation diversity. While diversity has become an important objective, existing methods still face fundamental limitations from both semantic and structural perspectives. On the one hand, multimodal recommender systems provide rich semantic signals, yet are often treated as auxiliary to ID embeddings, leading to ineffective semantic utilization for diversity. On the other hand, the accuracy--diversity trade-off persists during training, where accuracy gains often come at the cost of diversity. We show that this phenomenon is fundamentally governed by both the structural properties of learned representations and the dynamics of optimization. To address these challenges, we propose MOSS, a unified framework for multimodal decoupling through semantic and structural shaping. Specifically, we first introduce an MLLM-based Decoupled Feature Extraction module to disentangle multimodal content into informative representations. To enhance semantic utilization, we design a Category-aware Embedding Selection module that extracts compact, category-aware semantic features. To address structural limitations, we propose two complementary modules: Alternating Dual-Graph Propagation, which alleviates representation collapse by integrating collaborative and multimodal propagation, and Entropy-based Diversity Regularization, which mitigates optimization bias by promoting balanced category exposure. Extensive experiments on real-world datasets demonstrate that MOSS consistently achieves a superior accuracy--diversity trade-off over strong baselines.
\end{abstract}



\maketitle

\section{Introduction}

In the era of information overload, recommender systems play a crucial role in helping users efficiently discover relevant content \cite{raza2026comprehensive, hasan2025based, xu2026survey}. However, optimizing recommendation mainly for accuracy often leads to over-concentrated results, where similar items are repeatedly exposed and users' broader interests are insufficiently explored \cite{imran2023refrs, zhao2025fairness, areeb2023filter}. To improve user experience, recommendation diversity, usually reflected by the coverage of different item attributes such as categories, has become an important objective, as it benefits novelty, long-tail exposure, and balanced discovery.


Despite its importance, existing methods still face fundamental limitations. In particular, these limitations can be understood from two complementary perspectives: semantic utilization and structural modeling.

\textbf{C1: \textit{Semantic} limitation in multimodal utilization for di- versity-oriented recommendation.} 
Multimodal recommender systems introduce rich semantic signals from item content, such as textual and visual information, enabling items to be characterized by their intrinsic properties \cite{liu2024multimodal, xu2026survey, pomo2025recommender, lopez2025survey}. This provides a more expressive and semantically grounded representation of items.

However, most existing methods remain centered on item identity (ID) embeddings, treating multimodal information as an auxiliary signal through concatenation \cite{wei2020GRCN,he2016vbpr}, addition \cite{yu2023multi, guo2024lgmrec, zhou2023TMFUN}, or related fusion operations. Under this paradigm, multimodal content mainly enhances ID-driven representations, rather than being exploited as a primary source of semantic signals.

In principle, directly modeling items from multimodal content is particularly appealing for diversity-oriented recommendation. Unlike ID embeddings, which are largely shaped by historical user--item interactions and thus easily dominated by collaborative patterns such as popularity and exposure bias, multimodal information is grounded in the intrinsic characteristics of items themselves. It therefore provides richer semantic cues for distinguishing items from different perspectives, offering a more natural basis for diversified recommendation.

Nevertheless, directly exploiting multimodal features remains highly nontrivial. Recent studies attempt to learn item representations directly from multimodal features \cite{zhang2024dual, zhou2025learning, zhang2025freelunch}, yet prior works show that using pre-trained multimodal representations does not necessarily improve recommendation performance and may even degrade it \cite{yuan2023wheregonext,zhou2025learning}, as further analyzed in Section~\ref{sec:analysis1}. These observations indicate that the key challenge lies not in the availability of multimodal semantics, but in how to effectively organize and utilize them for diversity.

\begin{figure}[htbp]
\includegraphics[width=8cm]{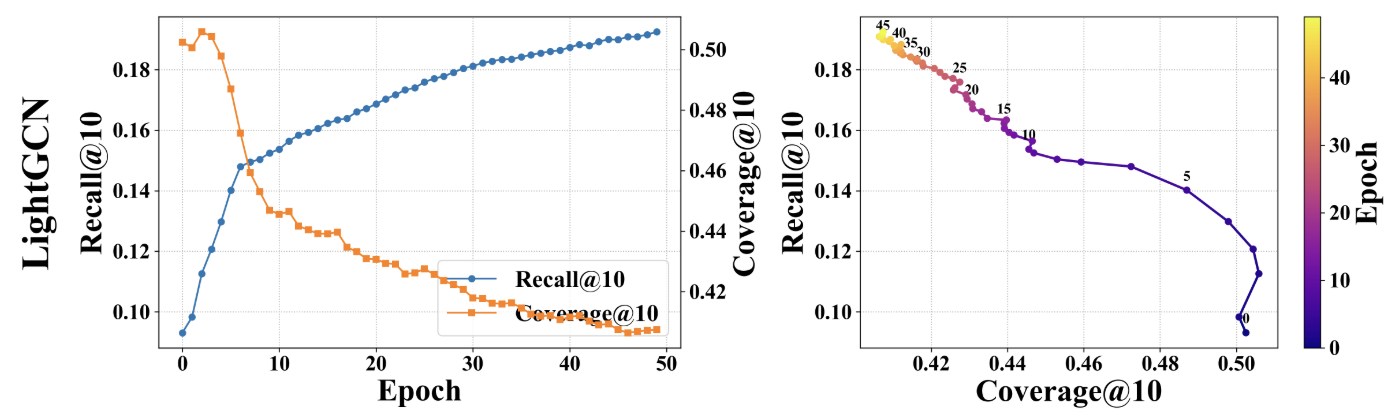}
\caption{Training dynamics and the accuracy--diversity trade-off of LightGCN on Anime, with Recall@10 for accuracy and Coverage@10 for diversity. The left figure shows Recall@10 and Coverage@10 over the first 50 epochs, and the right figure shows the corresponding Coverage@10--Recall@10 trajectory, colored by epoch and labeled every five epochs.}
\label{fig:tradeoff} 
\end{figure}

\textbf{C2: \textit{Structural} limitation in balancing accuracy and diversity during training.}
In collaborative recommendation, achieving high accuracy and high diversity simultaneously remains challenging. As illustrated in Fig.~\ref{fig:tradeoff}, this trade-off is observed in most GNN-based models, such as LightGCN \cite{he2020lightgcn}, where Recall@10 steadily improves during training, while Coverage@10 consistently declines. This indicates that accuracy gains are often achieved by concentrating recommendations on a narrower set of items, reflecting an increasing structural concentration in the learned representation space. As this pattern persistently emerges throughout training, it suggests that the accuracy--diversity trade-off is not merely an empirical phenomenon, but is closely related to the structural properties of learned \textit{representations} and the  \textit{optimization} dynamics. 

\textit{(1) Dimensional Collapse as a Representation-Level Bias.}
From the representation perspective, we relate this trade-off to dimensional collapse \cite{Liu2019DC} observed in GNN-based collaborative filtering \cite{chen2024DCinCF,ma2021unified,zhang2025DCinGCL}, where item representations are confined to a narrow subspace. Such structural compression reduces the effective capacity of the representation space, making items more similar and less distinguishable. As a result, the model becomes less capable of retrieving semantically diverse items, leading to homogeneous ranking outputs and reduced structural diversity. A more detailed analysis is provided in Section~\ref{sec:analysis2}.

\textit{(2) Accuracy-Oriented Objective as an Optimization-Level Bias.}
From the optimization perspective, the conventional recommendation objectives, such as BPR \cite{Rendle2009BPR}, further reinforce this structural concentration. By prioritizing ranking accuracy without explicit diversity constraints, these objectives favor items that consistently contribute to the optimization signal, typically popular items and dominant categories. Consequently, the representation space becomes increasingly biased toward a limited subset of items, resulting in improved accuracy but diminished diversity.

To address the above limitations, we propose a unified framework, termed \underline{\textbf{M}}ultim\underline{\textbf{o}}dal Decoupling for Diversified Recommendation
through \underline{\textbf{S}}emantic and \underline{\textbf{S}}tructural Shaping (MOSS), which consists of four complementary modules. First, we propose the \textit{MLLM-based Decoupled Feature Extraction} module, which leverages the reasoning capability of MLLMs to disentangle multimodal content and extract informative representations for the subsequent modules. 
To address the \textbf{first} challenge (C1), we propose the \textbf{\textit{semantic shaping}} module \textit{Category-aware Embedding Selection} module, which is introduced to tackle the underutilization of multimodal semantics in diversity-oriented recommendation. Instead of simply fusing modality information with ID embeddings or directly propagating raw multimodal features, it scores each embedding dimension from two complementary perspectives, namely global variance and category-level response balance, and retains a compact subset of informative and category-aware dimensions as the propagation input, thereby providing a stronger semantic basis for diversified recommendation. 
To address the \textbf{second} challenge (C2), we propose two \textbf{\textit{structural shaping}} modules. 
(1) \textit{Alternating Dual-Graph Propagation} module is designed to mitigate dimensional collapse, a representation-level bias that leads to diversity degradation in collaborative propagation. Specifically, it interleaves user--item collaborative propagation with multimodal semantic propagation over the item--item graph. This graph structural design alleviates dimensional collapse by enriching item representations with complementary multimodal signals, thereby preserving more expressive item relationships. Moreover, we theoretically show that this propagation scheme mitigates the tendency toward dimensional collapse. 
(2) \textit{Entropy-based Diversity Regularization} is designed to mitigate the optimization-level bias induced by accuracy-oriented objectives. Specifically, it imposes an entropy regularization term on the category distribution of high-ranked items, thereby reshaping the ranking structure toward a more balanced category distribution, thereby promoting category-level diversity while maintaining recommendation accuracy.

In summary, our main contributions are listed as follows:
\begin{itemize}
    \item To the best of our knowledge, this work presents the \textbf{first} unified framework that integrates multimodal recommendation and diversity modeling from both \textit{semantic} and \textit{structural} perspectives.
    
    \item From the \textit{semantic} perspective, we identify the underutilization of multimodal semantics in diversity-oriented recommendation. To address this limitation, we propose a \textit{Category-aware Embedding Selection} module, which extracts compact, category-aware, and semantically informative dimensions from decoupled multimodal representations, serving as a stable semantic anchor for downstream recommendation.

    \item From the \textit{structural} perspective, we analyze the accuracy--diversity trade-off as a consequence of both representation collapse and optimization bias, and propose a unified structural solution. Specifically, \textit{Alternating Dual-Graph Propagation} alleviates dimensional collapse through collaborative and multimodal propagation, enriching the structural expressiveness of item relationships. Furthermore, \textit{Entropy-based Diversity Regularization} mitigates optimization-induced bias by enforcing balanced category exposure, thereby improving diversity while preserving accuracy.

    \item Extensive experiments on two real-world datasets demonstrate that the proposed method consistently achieves a better accuracy--diversity trade-off than  baselines. 
\end{itemize}

\section{Analysis of Obstacles in Semantic and Structural Diversity Modeling}

To understand the limitations of existing methods in diversity-oriented recommendation, we analyze the problem from two complementary perspectives: semantic utilization and structural representation. The former focuses on how multimodal semantics are extracted and organized, while the latter examines how representations are distributed and evolved during training. 

\subsection{Semantic Limitation: Ineffective Utilization of Multimodal Semantics}
\label{sec:analysis1}



We begin by analyzing the role of multimodal semantics in recommendation. A natural approach is to directly initialize item representations with pre-trained multimodal features to enrich semantic information \cite{yuan2023wheregonext}. However, such direct utilization does not necessarily improve recommendation performance.

\begin{table}[htbp]
\centering
\caption{Effect of different item initialization strategies on representation quality and recommendation performance on Anime dataset.}
\resizebox{0.95\columnwidth}{!}{
\begin{tabular}{lccc}
\toprule
\textbf{Features} & \textbf{Norm. Eff.Rank} & \textbf{R@10}  & \textbf{C@10} \\
\midrule
ID embedding        & 0.99 & 0.1924 & 0.4074 \\
Visual embedding    & 0.65 & 0.1710 & 0.4032 \\
Textual embedding   & 0.68 & 0.1698 & 0.4096 \\
\bottomrule
\end{tabular}
}
\label{tab:analysis_erank}
\end{table}

As shown in Table~\ref{tab:analysis_erank}, replacing ID embeddings with visual or textual features under the same LightGCN backbone leads to a clear drop in Recall@10 (R@10), while Coverage@10 (C@10) does not show consistent improvement.

To further investigate this phenomenon, we analyze the normalized effective rank (Norm. Eff. Rank) \cite{Roy2007effectiverank}. Compared with ID-based initialization, multimodal initialization yields significantly lower effective rank, indicating that semantic information is unevenly distributed across embedding dimensions. This suggests that multimodal semantics are not effectively organized into discriminative representations, and may exhibit an early tendency toward representation collapse. To better understand this tendency, we further examine dimensional collapse in the following subsection.

As a result, although multimodal features contain rich semantic information, their effectiveness is not fully realized, limiting their contribution to diversity. These findings indicate that the key challenge lies not in the availability of multimodal semantics, but in how to organize them into compact and informative representations. This motivates the need for \textbf{\textit{semantic shaping}} to extract stable and category-aware features for downstream recommendation.


\begin{figure}[bp]
\includegraphics[width=8cm]{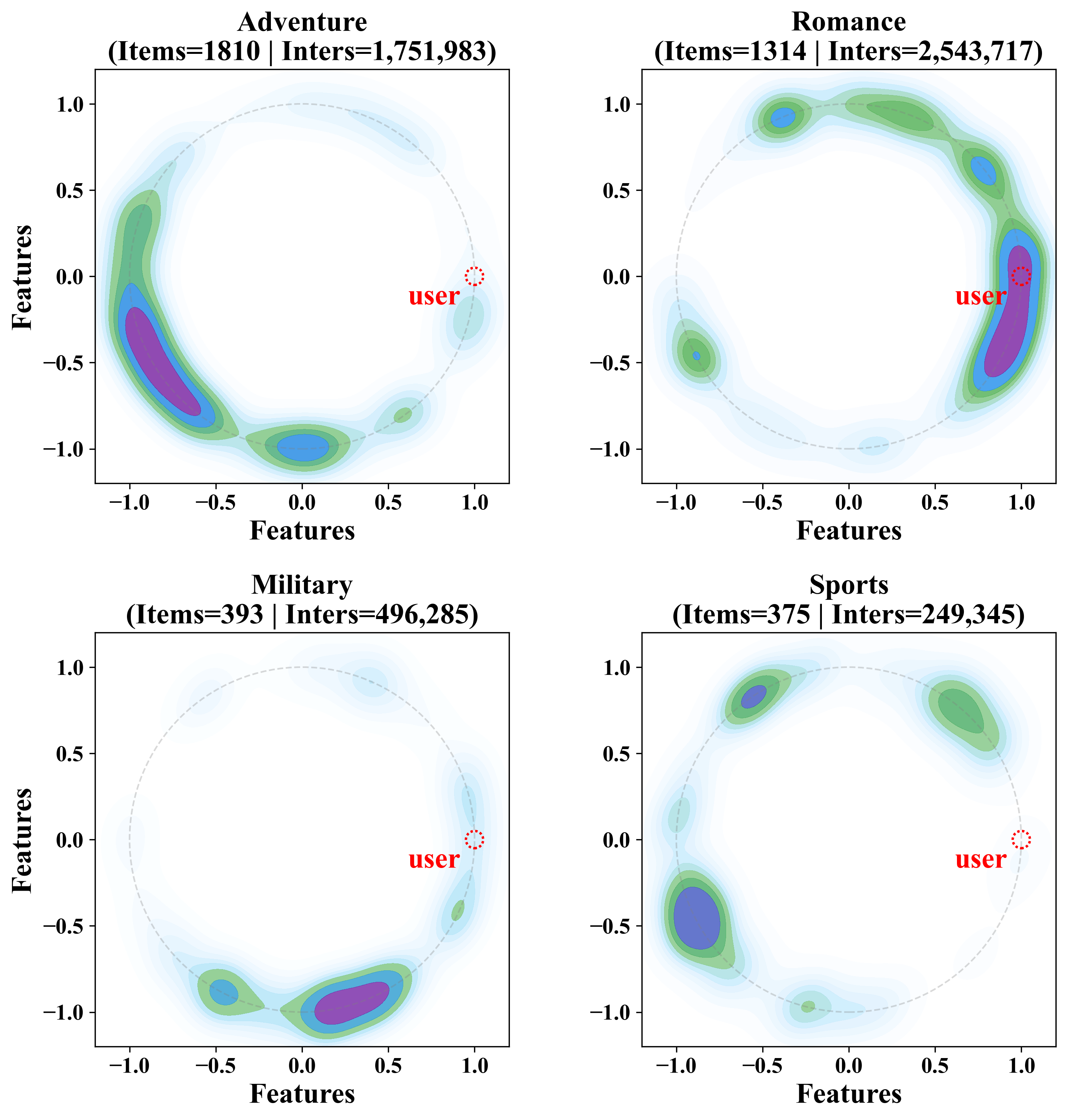}
\caption{Category-wise density visualization of item embeddings for four representative genres on Anime dataset. Embeddings are projected into 2D by t-SNE and normalized onto the unit circle, with filled contours represent density.}
\label{fig:categoryDC}
\end{figure}
\subsection{Structural Limitation: Dimensional Collapse and Diversity Degradation}

\label{sec:analysis2}

Building upon the above observations, we further examine how structural properties of representations affect diversity. Specifically, we analyze the distribution of item embeddings in the latent space \cite{ma2026SGMT,Wang2020UA,Zhang2023logdet,He2023CGCL}.

As illustrated in Fig.~\ref{fig:categoryDC}, item embeddings exhibit a clear dimensional collapse phenomenon, where representations are highly concentrated within each category with limited intra-category variation. While some degree of clustering is expected, the observed distributions reveal an excessive concentration tendency. Such structural compression reduces the discriminability among items and restricts the model’s ability to retrieve diverse candidates.

Consequently, recommendation results become increasingly homogeneous. For instance, users located near a dense region in the latent space are likely to receive recommendations from a narrow subset of items within the same category, while other categories remain underexplored.

From this perspective, the accuracy--diversity trade-off can be understood as a consequence of structural bias. Dimensional collapse reduces representation diversity, while accuracy-oriented objectives further reinforce concentration on dominant items and categories. Together, these effects lead to improved accuracy at the cost of diminished diversity.

This analysis highlights the necessity of \textbf{\textit{structural shaping}} to alleviate dimensional collapse while preserving meaningful semantic relationships, thereby enabling a better balance between accuracy and diversity.

\section{Methods}

\subsection{Preliminaries}

Let $\mathcal{U}$ and $\mathcal{I}$ denote the sets of users and items, respectively.
The interactions can be denoted by a matrix $\mathbf{R} \in \mathbb{R}^{n \times m}$, where $r_{ui} = 1$ if user $u$ interacts with item $i$, and $0$ otherwise.

In addition, each item may be associated with one or more category labels. 
Let $\mathcal{C}=\{c_1,\dots,c_{|\mathcal{C}|}\}$ denote the category set. 
For each item $i\in\mathcal{I}$, its category information is represented by a multi-hot category vector
$\mathbf{c}_i\in\{0,1\}^{|\mathcal{C}|}$, where the $k$-th entry equals 1 if item $i$ belongs to category $c_k$, and 0 otherwise.

\subsection{Overview}

\begin{figure*}[t]
    \centering
    \includegraphics[width=\textwidth]{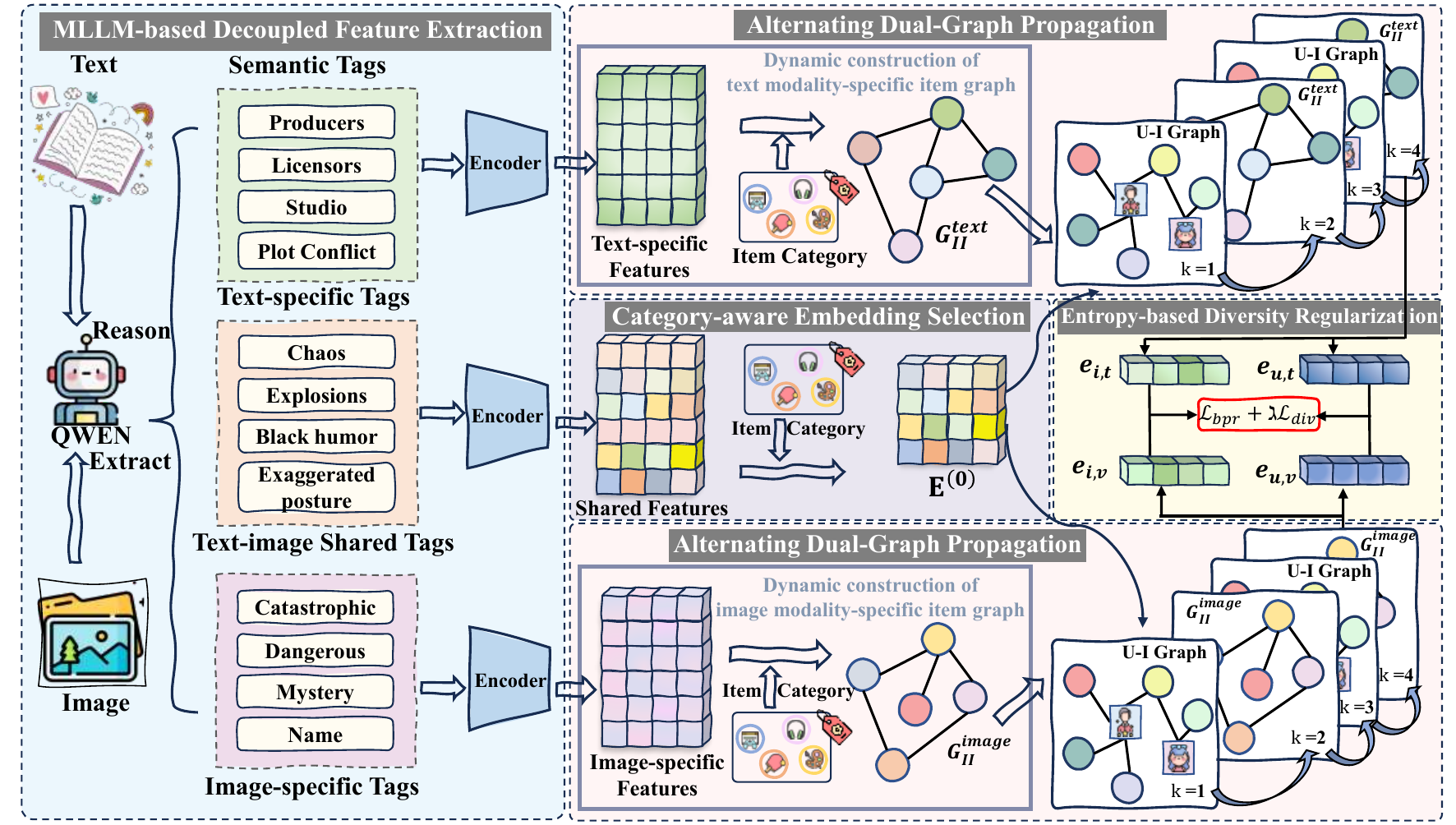}
    \caption{An illustration of the proposed MOSS model. The \textit{MLLM-based Decoupled Feature Extraction} module disentangles multimodal content into informative representations. The \textit{Category-aware Embedding Selection} module selects compact, category-aware features to enhance semantic utilization. The \textit{Alternating Dual-Graph Propagation} module alleviates dimensional collapse via collaborative and multimodal propagation. The \textit{Entropy-based Diversity Regularization} module promotes balanced category exposure while preserving accuracy.}
    \label{fig:framework}
\end{figure*}

Fig.~\ref{fig:framework} illustrates the overall architecture of MOSS, comprising of four core modules. For each item, we first employ an \textit{MLLM-based Decoupled Feature Extraction} to disentangle multimodal information into three complementary descriptions, namely shared, visual-specific, and textual-specific representations. 

Among them, the shared embedding is further refined by the proposed \textit{Category-aware Embedding Selection}, while the visual-specific and textual-specific embeddings are used to construct modality-specific item graphs.

Based on these representations, we design an \textit{Alternating Dual-Graph Propagation} module with two modality-specific branches. 
In each branch, propagation alternates between the user--item interaction graph and the corresponding category-modulated item graph, thereby progressively integrating collaborative, multimodal, and category-aware neighborhood information. 

Finally, we introduce an \textit{Entropy-based Diversity Regularization} term to explicitly encourage category-level diversity in the optimization level.


\subsection{MLLM-based Decoupled Feature Extraction}

In multimodal recommendation, items are often accompanied by visual and textual content, which offers rich and complementary semantic information beyond interaction signals.

To better exploit such multimodal semantics, we employ a Multimodal Large Language Model (MLLM), implemented with Qwen2.5-VL-72B \cite{bai2025qwen}, to decompose the content of item $i$ into three complementary components: a shared description $s_i$, a visual-specific description $v_i$, and a textual-specific description $t_i$.

These descriptions are then fed into the encoder, implemented by Sentence-BERT \cite{reimers2019sentencebert}, to obtain the shared, visual-specific, and textual-specific item representations ${h}_i^{s}$, ${h}_i^{v}$ and ${h}_i^{t}$. We denote the corresponding embedding matrices for all items by
$\mathbf{H}^{s}, \mathbf{H}^{v}, \mathbf{H}^{t}\in\mathbb{R}^{|\mathcal{I}|\times D}$, where $D$ denotes the original embedding dimension.

The shared embedding captures the common and dominant semantics of each item, and is used as the semantic anchor for graph propagation. 
By contrast, the modality-specific embeddings preserve view-specific semantic cues and are used to construct modality-specific item graphs, enabling the model to explore broader yet semantically related neighborhoods from different views.

\subsection{Category-aware Embedding Selection}
\label{sec:category_selection}

\begin{figure}[htbp]
\includegraphics[width=8.1cm]{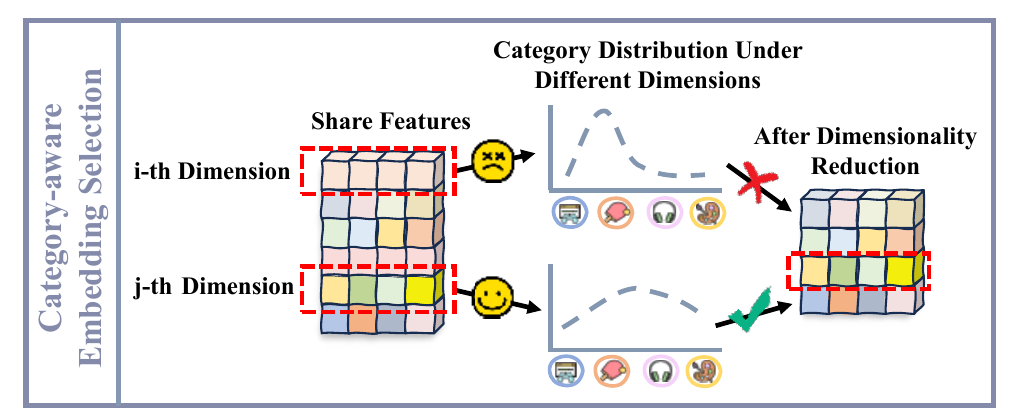}
\caption{Category-aware Embedding Selection.}
\label{fig:CAES}
\end{figure}

As illustrated in Fig.~\ref{fig:CAES}, given the rich semantics encoded in multimodal item embeddings, we expect the initialized embeddings for graph propagation to preserve informative multimodal cues for recommendation accuracy while promoting cross-category semantic sharing rather than overemphasizing category-exclusive patterns. 

As the shared embedding serves as the propagation input, this module is applied to the shared embedding matrix $\mathbf{H}^{s}\in\mathbb{R}^{|\mathcal{I}|\times D}$.

We first estimate the informativeness of each dimension by its global variance across all items:
\begin{equation}
\mathrm{Var}_j = \frac{1}{|\mathcal{I}|}\sum_{i=1}^{|\mathcal{I}|}\left(h_{ij}^{s}-\mu_j\right)^2,
\qquad
\mu_j=\frac{1}{|\mathcal{I}|}\sum_{i=1}^{|\mathcal{I}|}h_{ij}^{s},
\end{equation}
where $h_{ij}^{s}$ denotes the $j$-th dimension of the embedding of item $i$, and $\mu_j$ is the corresponding dataset-level mean. We then normalize the resulting variance into $v_j\in[0,1]$ by min--max scaling.

Variance serves as a simple indicator of how actively a dimension is utilized in the embedding space. Dimensions with larger variance better capture inter-item differences and are thus more likely to preserve informative multimodal semantics, whereas low-variance dimensions tend to be weakly activated across items, reflecting underutilized or partially collapsed dimensions.

Beyond variance, to quantify whether a feature dimension is shared across categories or dominated by only a few of them, we compute the average response magnitude of dimension $j$ within each category:
\begin{equation}
r_{c,j} = \frac{1}{|\mathcal{I}_c|}\sum_{i\in\mathcal{I}_c}|h_{ij}^{s}|,
\end{equation}
where $\mathcal{I}_c$ denote the set of items belonging to category $c$, with $r_{c,j}$ reflects how strongly the $j$-th dimension responds to category $c$ on average. 

Based on $\{r_{c,j}\}_{c=1}^{C}$, we construct a category-wise response distribution for dimension $j$:
\begin{equation}
p_{c,j} = \frac{r_{c,j}}{\sum_{c'=1}^{C}r_{c',j}}.
\end{equation}
A more concentrated $\{p_{c,j}\}_{c=1}^{C}$ indicates stronger category dominance, whereas a flatter distribution suggests more balanced cross-category responses.

We use the normalized entropy to measure the category-level balance of dimension $j$:
\begin{equation}
H_j = -\frac{1}{\log C}\sum_{c=1}^{C} p_{c,j}\log\left(p_{c,j}\right).
\end{equation}
A larger $H_j$ indicates that the responses of this dimension are distributed more evenly across categories, and are more likely to capture shared semantics rather than category-specific bias.

Finally, we combine global informativeness and category response balance to score the $j$-th dimension, with $\alpha\in[0,1]$ balancing informativeness and category response balance:
\begin{equation}
s_j = (1-\alpha) v_j + \alpha H_j,
\label{eq:cat_selection_score}
\end{equation}

We then retain the top-$d$ dimensions with the highest scores. For each item $i$, the selected shared embedding is defined as
\begin{equation}
\mathbf{x}_{i}
=
\big[h^{s}_{i,j}\mid s_j \in \mathrm{top}\text{-}d(\{s_j\}_{j=1}^{D})\big],
\end{equation}
where $h^{s}_{i,j}$ denotes the $j$-th entry of the shared embedding for item $i$. By stacking $\mathbf{x}_i$ for all items, we obtain the selected shared embedding matrix $\mathbf{X}\in\mathbb{R}^{|\mathcal{I}|\times d}$, which is used as the initialized item-side input for subsequent graph propagation.



\subsection{Alternating Dual-Graph Propagation}
\label{sec:ADGP}


Starting from the selected shared embeddings $\mathbf{X} \in \mathbb{R}^{|\mathcal{I}| \times d}$, we perform graph propagation to jointly exploit collaborative structure and modality-specific semantic relations. Specifically, the selected shared embeddings serve as the item-side initialization for propagation, providing a stable semantic anchor shared across branches, while the modality-specific embeddings are used to identify semantically related neighbors from the perspective of each modality. In addition, category information is incorporated to further refine the graph structure, enabling the model to capture more informative and diverse item relations. 

Let $\mathbf{E}_u\in\mathbb{R}^{|\mathcal{U}|\times d}$ denote the trainable user embedding matrix, and let $\mathbf{e}_u^{(0)}$ be the initial embedding of user $u$. For each modality branch $m\in\{t,v\}$, the initial representations are defined as
\begin{equation}
{(\mathbf{e}_{u}^m)}^{(0)}=\mathbf{e}_u^{(0)}, \qquad
{(\mathbf{e}_{i}^m)}^{(0)}=\mathbf{x}_i.
\end{equation}

For each modality $m\in\{t,v\}$, we construct a category-modulated item graph from corresponding modality-specific features. 

Given the modality-specific feature $\mathbf{h}_i^{m}$ and the category vector $\mathbf{c}_i$ of item $i$, we compute the modality similarity and category similarity as
\begin{equation}
s_{ij}^{m}
=
\frac{(\mathbf{h}_i^{m})^{\top}\mathbf{h}_j^{m}}
{\|\mathbf{h}_i^{m}\|_2\|\mathbf{h}_j^{m}\|_2},
\qquad
s_{ij}^{c}
=
\frac{\mathbf{c}_i^{\top}\mathbf{c}_j}
{\|\mathbf{c}_i\|_2\|\mathbf{c}_j\|_2},
\end{equation}
and define the fused similarity by
\begin{equation}
\tilde{s}_{ij}^{m}=s_{ij}^{m}\cdot s_{ij}^{c}.
\end{equation}

For each item $i$, we first retain the top-$K_c$ items under $\tilde{s}_{ij}^{m}$ to form a candidate pool $\mathcal{P}_i^{m}$, and then randomly sample $K_s$ neighbors at each propagation layer:
\begin{equation}
\mathcal{P}_i^{m}=\left\{j \;\middle|\; \tilde{s}_{ij}^{m}\in \mathrm{top}\text{-}K_c\!\left(\{\tilde{s}_{ij}^{m}\}\right)\right\}, \qquad
\mathcal{N}_i^{m}\subseteq \mathcal{P}_i^{m},\ |\mathcal{N}_i^{m}|=K_s.
\end{equation}
Here, $K_c$ and $K_s$ denote the candidate pool size and the number of sampled neighbors, and $\mathcal{N}_i^{m}$ denotes the sampled neighbor set used in the current propagation layer. In this way, $\mathcal{P}_i^{m}$ preserves the most relevant modality-aware neighbors, while random sampling enables different yet relevant local structures to be explored across layers and training iterations, thereby enhancing structural diversity and representation learning.

The corresponding weighted adjacency matrix is then defined as
\begin{equation}
(\mathbf{A}_{ii}^{m})_{ij}
=
\begin{cases}
\tilde{s}_{ij}^{m}, & j\in\mathcal{N}_i^{m},\\
0, & \text{otherwise}.
\end{cases}
\end{equation}
We then normalize it as $\hat{\mathbf{A}}_{ii}^{m}$. For simplicity, let $a_{ij}^{m}=(\hat{\mathbf{A}}_{ii}^{m})_{ij}$ denote the normalized edge weight between items $i$ and $j$ in branch $m$.

To incorporate collaborative signal and enrich structural diversity, we adopt an alternating propagation mechanism over the user--item graph and the modality-specific item graph. 
At layer $k$, let ${(\mathbf{e}_{u}^m)}^{(k)}$ and ${(\mathbf{e}_{i}^m)}^{(k)}$ denote the representations of user $u$ and item $i$ in branch $m$, respectively. 
$\mathcal{N}_{u}$ and $\mathcal{N}_{i}$ represent the set of items or users connected to user $u$ or item $i$ in the user--item graph.

For even-numbered layers, collaborative propagation is performed on the normalized user--item graph:
\begin{equation}
{(\mathbf{e}_{u}^{m})}^{(k+1)}
=
\sum_{i\in\mathcal{N}_{u}}
\frac{1}{\sqrt{|\mathcal{N}_{u}|}\sqrt{|\mathcal{N}_{i}|}}
{(\mathbf{e}_{i}^m)}^{(k)},
\end{equation}
\begin{equation}
{(\mathbf{e}_{i}^{m})}^{(k+1)}
=
\sum_{u\in\mathcal{N}_{i}}
\frac{1}{\sqrt{|\mathcal{N}_{i}|}\sqrt{|\mathcal{N}_{u}|}}
{(\mathbf{e}_{u}^{m})}^{(k)}.
\end{equation}

For odd-numbered layers, the user representations remain unchanged, while the item representations are updated on the current modality-specific item graph:
\begin{equation}
{(\mathbf{e}_{u}^m)}^{(k+1)}={(\mathbf{e}_{u}^m)}^{(k)}, \quad
{(\mathbf{e}_{i}^m)}^{(k+1)}
=
\sum_{j\in\mathcal{N}_{i}^{m}}
a_{ij}^{m}\,{(\mathbf{e}_{j}^m)}^{(k)}.
\end{equation}

Collaborative propagation captures user preference from interaction signal while item-graph propagation introduces related neighbors from modality-specific perspectives, leading to more informative and diversity-aware item representations.

After $L$ propagation layers, we obtain the final representations by mean pooling over all layers:
\begin{equation}
\mathbf{e}_{u}^m
=
\frac{1}{L+1}\sum_{k=0}^{L}{(\mathbf{e}_{u}^m)}^{(k)},
\quad
\mathbf{e}_{i}^m
=
\frac{1}{L+1}\sum_{k=0}^{L}{(\mathbf{e}_{i}^m)}^{(k)}.
\end{equation}

This layer-wise aggregation helps preserve information from different propagation depths, and thus mitigates the loss of useful signals caused by excessive smoothing. 
Thus, we obtain two propagated outputs, i.e., $\{\mathbf{e}_{u}^t,\mathbf{e}_{i}^t\}$ and $\{\mathbf{e}_{u}^v,\mathbf{e}_{i}^v\}$. 
We then fuse the two modality branches by simple concatenation:
\begin{equation}
\mathbf{e}_u = \mathbf{e}_{u}^t \,\|\, \mathbf{e}_{u}^v,
\qquad
\mathbf{e}_i = \mathbf{e}_{i}^t \,\|\, \mathbf{e}_{i}^v,
\end{equation}
where $\|$ denotes concatenation, with $\mathbf{e}_u$ and $\mathbf{e}_i$ represents the user and item representations produced by the propagation module.

\subsection{Entropy-based Diversity Regularization }

To further encourage recommendation diversity, we introduce an entropy-based regularizer to encourage category-level diversity.

For each user $u$, given the predicted score $\hat{y}_{ui}=\mathbf{e}_u^\top \mathbf{e}_i$ for item $i$, we normalize the category vector $\mathbf{c}_i$ as $\mathbf{p}_i=\frac{\mathbf{c}_i}{\|\mathbf{c}_i\|_1}$, where $\mathbf{p}_i\in\mathbb{R}^C$ denotes the normalized category distribution of item $i$, and $\|\mathbf{c}_i\|_1$ is the number of category labels of item $i$. We then define a soft importance distribution:
\begin{equation}
w_{ui}
=
\frac{\exp(\hat{y}_{ui})}
{\sum_{j\in\mathcal{I}}\exp(\hat{y}_{uj})},
\qquad i\in\mathcal{I},
\end{equation}
where $w_{ui}$ denotes the relative contribution of item $i$ to the recommendation distribution of user $u$. For efficiency, this soft distribution is approximated over the top-20 scored items in practice.

Using these score-aware weights, we construct the user-specific category distribution as
\begin{equation}
\mathbf{q}_u
=
\sum_{i \in \mathcal{I}} w_{ui}\,\mathbf{p}_i.
\end{equation}
Here, $\mathbf{q}_u \in \mathbb{R}^{C}$ can be regarded as the expected category composition induced by the predicted ranking scores of user $u$, and its $c$-th entry is denoted by $q_{u,c}$.

To encourage the recommendation results to cover a broader range of categories, we maximize the entropy of $\mathbf{q}_u$. 
Equivalently, we minimize its negative entropy:
\begin{equation}
\mathcal{L}_{\mathrm{div}}
=
\frac{1}{|\mathcal{B}|}
\sum_{u \in \mathcal{B}}
\sum_{c=1}^{C}
q_{u,c}\log(q_{u,c}),
\end{equation}
where $\mathcal{B}$ is the user mini-batch. 

This regularizer assigns a larger penalty when the predicted recommendation distribution is overly concentrated on only a few categories, and a smaller penalty when the category distribution is more balanced. 
Therefore, it complements the ranking objective by explicitly promoting category-level diversity in the final recommendation outputs.

The overall training objective is defined as
\begin{equation}
\mathcal{L}
=
\mathcal{L}_{\mathrm{BPR}}
+
\lambda_{\mathrm{div}}\,\mathcal{L}_{\mathrm{div}}
+
\lambda_{reg}\,\lVert \Theta \rVert_2.
\end{equation}
where $\mathcal{L}_{\mathrm{BPR}}$ is the pairwise ranking loss, $\Theta$ is the set of learned user and item embeddings, $\lambda_{\mathrm{div}}$ controls the strength of entropy-based diversity regularization and $\lambda_{reg}$ is the regularization coefficient.

\section{Theoretical Analysis of the Mitigation of Dimensional Collapse}
\label{sec:theory}
To theoretically show how \textit{Alternating Dual-Graph Propagation} may mitigate the issue of dimensional collapse. We analyze dimensional collapse from the perspective of spectral concentration after propagation. 
Let the two-layer propagated representation be
\begin{equation}
E^{(2)} = Q E^{(0)},
\end{equation}
where $E^{(0)} \in \mathbb{R}^{(|\mathcal U|+|\mathcal I|)\times d}$ is the initial representation matrix and $Q$ denotes the corresponding two-layer propagation operator.

Dimensional collapse means that the propagated representation becomes increasingly dominated by only a few principal directions, while the remaining directions are severely compressed. 
To characterize this phenomenon, we consider the Gram matrix of the propagated representation:
\begin{equation}
G^{(2)} = (E^{(2)})^\top E^{(2)}= (E^{(0)})^\top Q^\top Q E^{(0)}.
\label{eq:gram}
\end{equation}
A more concentrated spectrum of $G^{(2)}$ implies that most representation energy lies in only a few dominant directions, indicating a stronger dimensional collapse tendency.

Equation.~\ref{eq:gram} shows that the spectral concentration of the propagated representation is primarily governed by the operator $Q^\top Q$. Since $Q^\top Q$ is positive semidefinite, its eigenvalues satisfy
\begin{equation}
\lambda_j(Q^\top Q)=\sigma_j(Q)^2,
\end{equation}
where $\sigma_j(Q)$ denotes the $j$-th singular value of $Q$. Therefore, the singular value spectrum of $Q$ directly determines how propagation redistributes representation energy across different directions. 

Following the above spectral characterization, we specifically consider two two-layer propagation operators for comparison:
\begin{equation}
Q_{2UI}=P_{ui}^2,\qquad Q_{UI+II}=P_{ii}P_{ui},
\end{equation}
where
\begin{equation}
P_{ui}=
\begin{bmatrix}
0 & R\\
R^\top & 0
\end{bmatrix},
\qquad
P_{ii}=
\begin{bmatrix}
I & 0\\
0 & S
\end{bmatrix}.
\end{equation}

Here, $R$ denotes the normalized user--item biadjacency matrix, $S$ denotes the normalized item--item propagation matrix, and $I$ is the identity matrix on the user side. The construction of $S$ is consistent with Section~\ref{sec:ADGP}. For theoretical and empirical simplicity, we adopt a static top-10 item--item graph, ignoring the dynamic update strategy used in the full model.  $Q_{2UI}$ corresponds to pure U--I propagation, while $Q_{UI+II}$ incorporates item--item propagation.

\begin{figure}[htbp]
\includegraphics[width=6cm]{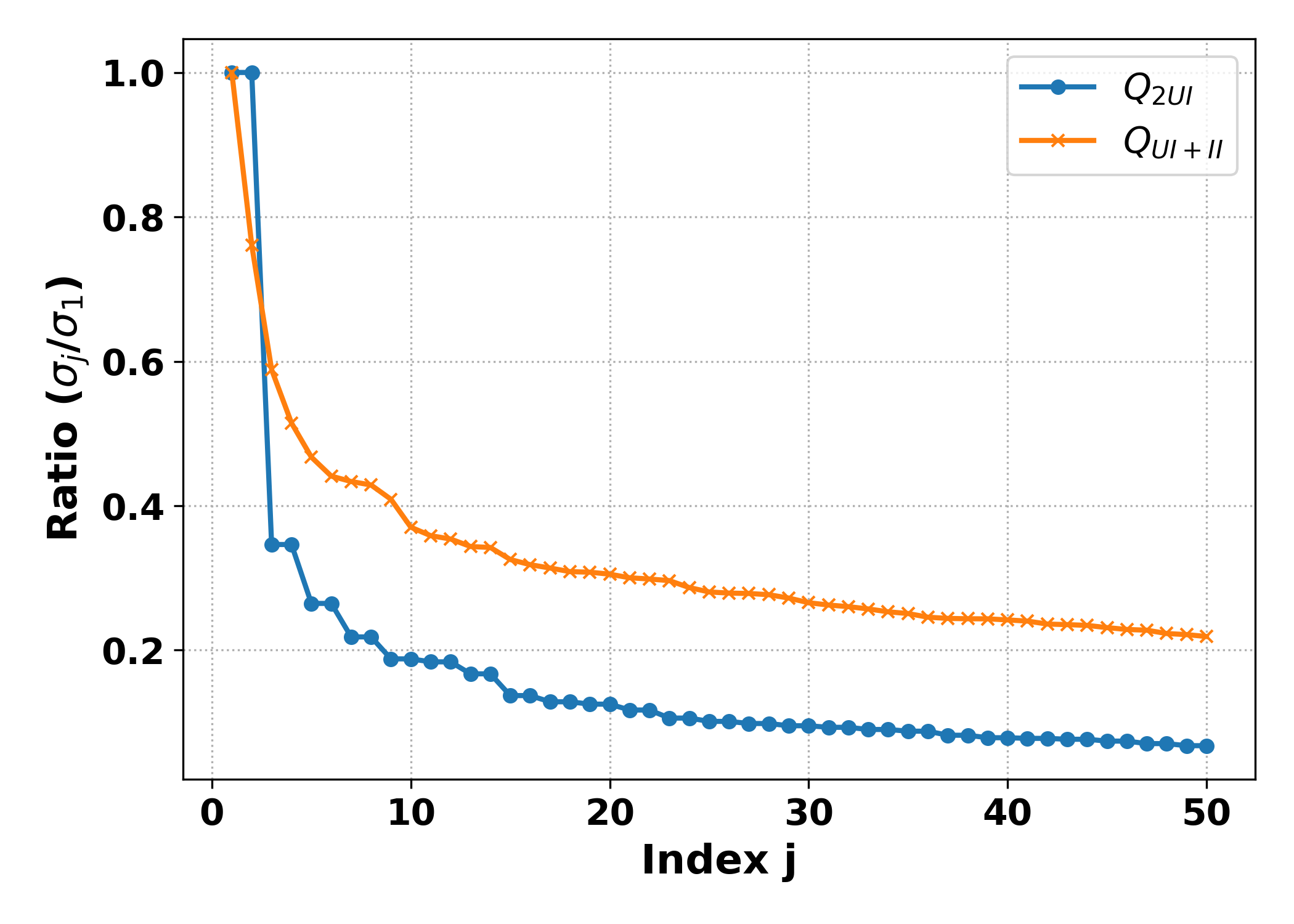}
\caption{Comparison of the top-$50$ normalized singular value of two two-layer propagation operators on Anime. The curves show the normalized singular value ratios $\sigma_j/\sigma_1$ for $Q_{2UI}$ and $Q_{UI+II}$, sorted by singular value magnitude.}
\label{fig:SVcompareV}
\end{figure}

As shown in Fig.~\ref{fig:SVcompareV}, the normalized singular value ratio curve of $Q_{UI+II}$ is consistently flatter than that of $Q_{2UI}$, with higher $\sigma_j/\sigma_1$ for most non-leading components. Therefore, the energy redistribution induced by $Q_{UI+II}$ is less concentrated on only a few principal directions, leading to a weaker concentration tendency in the propagated Gram matrix $G^{(2)}_{UI+II}$. This indicates that alternating UI--II propagation preserves substantially stronger non-leading components, whereas pure two-layer collaborative propagation over the user--item graph suppresses them much more aggressively. Consequently, from the operator perspective, introducing item--item propagation helps alleviate the dimensional collapse tendency inherent in collaborative propagation.

\section{Experiments}
\subsection{Experimental Setup}
\subsubsection{Datasets} Experiments are conducted on two datasets, Anime \footnote{\url{https://www.kaggle.com/datasets/hernan4444/anime-recommendation-database-2020}} \cite{girsang2020collaborative, putri2023analyzing} and Game \footnote{\url{https://github.com/kang205/SASRec}} \cite{wan2018item, kang2018self, pathak2017generating}, using a 5-core setting that retains users and items with a minimum of five interactions. Summary statistics are provided in Table~\ref{table_dataset}. 
\begin{table}[h] \centering \setlength{\abovecaptionskip}{0.1cm} \setlength{\belowcaptionskip}{-0.1cm} \caption{Statistics of datasets.} \label{table_dataset} 
\begin{subtable}{0.45\linewidth} \centering \setlength{\belowcaptionskip}{-0.1cm} \caption{Game Dataset} 
\label{table:game_dataset} 
\begin{tabular}{lr} \toprule \# Players & 195,184 \\ \# Games & 10,425 \\ \# Interactions & 2,294,883 \\ \# Categories & 22 \\ \bottomrule \end{tabular} \end{subtable} \hfill 
\begin{subtable}{0.45\linewidth} \centering \setlength{\belowcaptionskip}{-0.1cm} \caption{Anime Dataset} 
\label{table:anime_dataset} \begin{tabular}{lr} \toprule \# Players & 66,870 \\ \# Animes & 8,519 \\ \# Interactions & 7,793,920 \\ \# Categories & 45 \\ \bottomrule \end{tabular} \end{subtable} \end{table}
\begin{table*}[t]
\centering
\caption{Comparison of models on the Anime and Game datasets.}
\label{tab:overall}

\small
\setlength{\tabcolsep}{2.6pt}
\renewcommand{\arraystretch}{1.0}

\setlength{\heavyrulewidth}{0.08em}
\setlength{\lightrulewidth}{0.05em}
\setlength{\cmidrulewidth}{0.03em}
\setlength{\arrayrulewidth}{0.4pt}

\begin{tabular*}{\textwidth}{
@{\extracolsep{\fill}}
cc|ccc|cccc|cc
@{}
}
\toprule

\multirow[c]{2}{*}{\textbf{Dataset}}
& \multirow[c]{2}{*}{\textbf{Metric}}
& \multicolumn{3}{c|}{\textbf{Accuracy-based Methods}}
& \multicolumn{4}{c|}{\textbf{Diversity-based Methods}}
& \multicolumn{2}{c}{\textbf{Our Method}}
\\

\cmidrule(lr){3-5}
\cmidrule(lr){6-9}
\cmidrule(lr){10-11}

&
& LightGCN
& SimGCL
& Freedom
& DGCN
& DGRec
& DRGame
& DivGCL
& MOSS
& MOSS$_{\mathrm{acc}}$
\\

\midrule

\multirow[c]{5}{*}{\textbf{Anime}}
& Recall@10
& 0.1924
& 0.2100
& 0.1985
& 0.1106
& 0.2024
& 0.0569
& \underline{0.2132}
& 0.1860
& \textbf{0.2211}
\\

& NDCG@10
& 0.2124
& 0.2248
& 0.2014
& 0.1169
& 0.1868
& 0.0975
& \underline{0.2329}
& 0.1756
& \textbf{0.2374}
\\

& Coverage@10
& 0.4074
& 0.4387
& 0.4345
& \underline{0.4890}
& 0.4428
& 0.4279
& 0.4197
& \textbf{0.5536}
& 0.4417
\\

& Entropy@10
& 5.0054
& 5.7738
& 5.3682
& \underline{6.1741}
& 5.5891
& 5.6576
& 5.1921
& \textbf{6.9645}
& 6.0342
\\

& GM@10
& 0.2800
& 0.3035
& 0.2937
& 0.2326
& 0.2994
& 0.1560
& 0.2991
& \textbf{0.3209}
& \underline{0.3125}
\\

\midrule

\multirow[c]{5}{*}{\textbf{Game}}
& Recall@10
& 0.1080
& \underline{0.1109}
& 0.1095
& 0.0803
& 0.1079
& 0.0193
& 0.1108
& 0.0907
& \textbf{0.1114}
\\

& NDCG@10
& 0.0604
& 0.0614
& 0.0615
& 0.0452
& 0.0602
& 0.0069
& \underline{0.0617}
& 0.0391
& \textbf{0.0621}
\\

& Coverage@10
& 0.3371
& 0.3431
& 0.3298
& 0.3636
& 0.3552
& \underline{0.4872}
& 0.3544
& \textbf{0.6348}
& 0.3580
\\

& Entropy@10
& 2.0726
& 2.2041
& 2.2450
& 2.3932
& 2.2190
& \underline{3.3367}
& 2.1703
& \textbf{3.9764}
& 2.8757
\\

& GM@10
& 0.1908
& 0.1951
& 0.1900
& 0.1709
& 0.1958
& 0.0970
& 0.1981
& \textbf{0.2400}
& \underline{0.1997}
\\

\bottomrule
\end{tabular*}

\end{table*}

\begin{table*}[t]
\centering
\caption{Ablation study on the Anime and Game datasets.}
\label{tab:ablation_study}

\small
\setlength{\tabcolsep}{2.6pt}
\renewcommand{\arraystretch}{1.0}

\setlength{\heavyrulewidth}{0.08em}
\setlength{\lightrulewidth}{0.05em}
\setlength{\cmidrulewidth}{0.03em}
\setlength{\arrayrulewidth}{0.4pt}

\begin{tabular*}{\textwidth}{
@{\extracolsep{\fill}}
c|ccccc|ccccc
@{}
}
\toprule

&
\multicolumn{5}{c|}{\textbf{Anime}}
&
\multicolumn{5}{c}{\textbf{Game}}
\\

\cmidrule(lr){2-6}
\cmidrule(lr){7-11}

\textbf{Model}
& Recall@10
& NDCG@10
& Coverage@10
& Entropy@10
& GM@10
& Recall@10
& NDCG@10
& Coverage@10
& Entropy@10
& GM@10
\\

\midrule

$w/o\ \mathrm{EBDR}$
& \textbf{0.2172}
& \textbf{0.2306}
& 0.4577
& 6.0258
& 0.3153
& \textbf{0.1092}
& \textbf{0.0607}
& 0.3642
& 2.9347
& 0.1994
\\

$w/o\ \mathrm{ADGP}$
& 0.1730
& 0.1619
& 0.5498
& 6.5569
& 0.3084
& 0.0863
& 0.0384
& 0.6127
& 3.8642
& 0.2299
\\

$w/o\ \mathrm{CAES}$
& 0.1722
& 0.1591
& 0.5187
& 6.5487
& 0.2989
& 0.0852
& 0.0372
& 0.5932
& 3.8533
& 0.2248
\\

MOSS
& 0.1860
& 0.1756
& \textbf{0.5536}
& \textbf{6.9645}
& \textbf{0.3209}
& 0.0907
& 0.0391
& \textbf{0.6348}
& \textbf{3.9764}
& \textbf{0.2400}
\\

\bottomrule
\end{tabular*}

\end{table*}



\subsubsection{Baselines and Evaluation Metrics}
We benchmark MOSS against two categories of baselines. For accuracy evaluation, we include classical graph collaborative filtering models, LightGCN \cite{he2020lightgcn} and SimGCL \cite{liu2024simgcl}, as well as the multimodal recommendation model Freedom \cite{zhou2023tale}. To assess diversity, we compare MOSS with several modality-agnostic diversity-enhanced methods, including DGCN \cite{dgcn}, DGRec \cite{dgrec}, DRGame \cite{liu2024drgame}, and DivGCL \cite{gong2025divgcl}. Top-$K$ recommendation performance is evaluated using four standard metrics: Coverage and Entropy for diversity, Recall and Normalized Discounted Cumulative Gain for accuracy, and $K$ is set to 10. To further quantify the balance between accuracy and diversity, we compute the geometric mean (GM) \cite{zhang2025dual} of Recall@$K$ and Coverage@$K$:
\begin{equation}
    GM@K = \sqrt{\mathrm{Recall}@K \times \mathrm{Coverage}@K}.
\end{equation}





\subsection{Overall Performance}
\subsubsection{Performance Evaluation}

\begin{figure}[htpb]
    \centering

    \begin{subfigure}{0.49\linewidth}
        \centering
        \includegraphics[width=\linewidth]{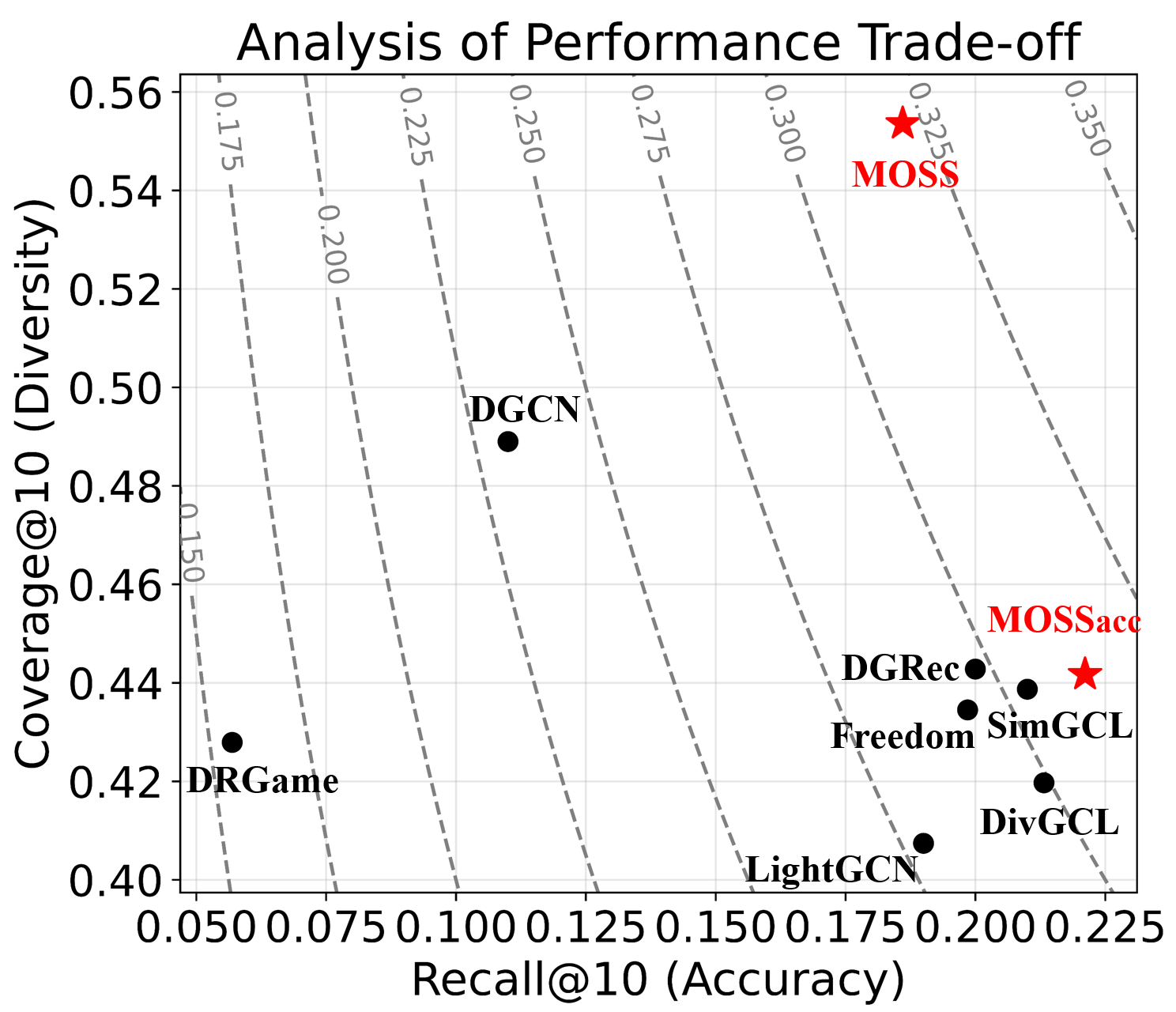}
        \caption{Anime dataset.} 
        \label{fig:GM-anime}
    \end{subfigure}
    \hfill
    \begin{subfigure}{0.49\linewidth}
        \centering
        \includegraphics[width=\linewidth]{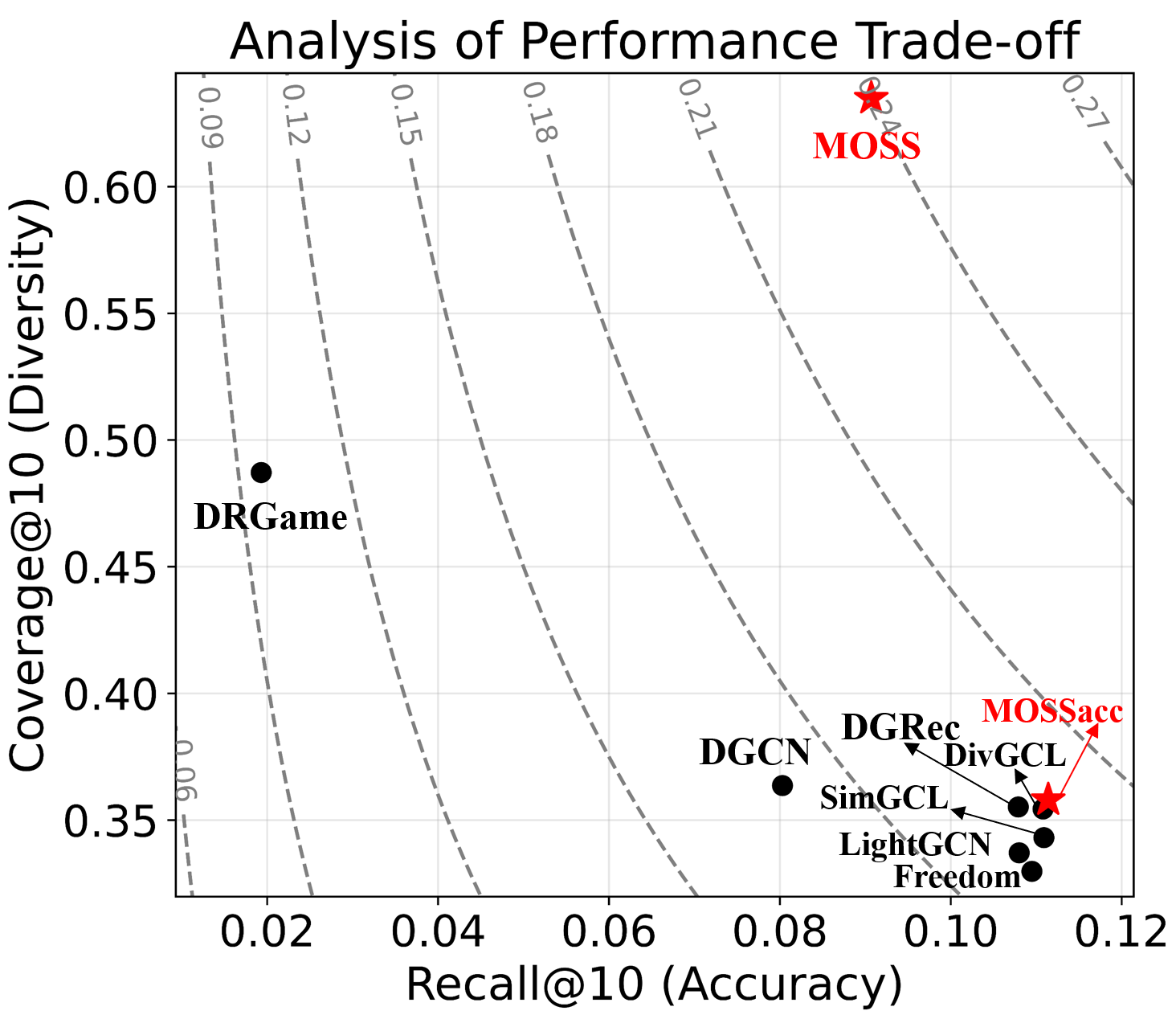}
        \caption{Game dataset.}
        \label{fig:GM-game}
    \end{subfigure}
    \caption{Comparison of the geometric mean (GM) between MOSS and baseline methods.}
    \label{fig:GM}
\end{figure}

We compare the proposed model \textbf{MOSS} with several representative baselines. The results are reported in Table~\ref{tab:overall}, where the best and second-best results are highlighted in \textbf{bold} and \underline{underlined}, respectively. Based on these results, we draw the following observations:
(1) \textbf{Superior diversity.} MOSS achieves the best diversity performance on both datasets, improving Coverage@10 / Entropy@10 by 13.2\% / 12.8\% on \textit{Anime} and 30.3\% / 19.2\% on \textit{Game} over the strongest diversity-oriented baseline.
(2) \textbf{Strong accuracy under accuracy-oriented settings.} Admittedly, the diversity-oriented version of MOSS is not always the most accurate, since it explicitly favors broader category coverage. However, when we tuning the relevant hyperparameters toward more accuracy-oriented settings, MOSS$_{\text{acc}}$ can also achieve the best accuracy on \textit{Anime} and \textit{Game}. This suggests that Modules \textit{Alternating Dual--Graph Propagation} and \textit{Category-aware Embedding Selection} are not only beneficial for diversity, but also improve recommendation accuracy by better exploiting multimodal semantics and structural signals.
(3) \textbf{Best accuracy--diversity trade-off.} As shown in Fig.~\ref{fig:GM}, MOSS achieves a more favorable accuracy--diversity balance than all baselines.  Improving GM@10 by 5.7\% and 21.1\% on Anime and Game.

\subsubsection{Ablation Study}

To evaluate the contribution of each key component, we perform ablation studies on both datasets. As shown in Table~\ref{tab:ablation_study}, ``w/o CAES'', ``w/o ADGP'', and ``w/o EBDR'' denote the removal of \textit{Category-aware Embedding Selection}, \textit{Alternating Dual-Graph Propagation}, and \textit{Entropy-based Diversity Regularization}, respectively. The results confirm that each component provides measurable gains. Specifically, EBDR primarily promotes diversity, while both ADGP and CAES contribute to improvements in both accuracy and diversity.
\section{Conclusion}
In this paper, we study diversified recommendation from both semantic and structural perspectives, showing that multimodal semantics are underutilized and that the accuracy--diversity trade-off is driven by representation collapse and optimization bias. We propose \textbf{MOSS}, a unified framework combining multimodal decoupling with semantic and structural shaping, including feature extraction, category-aware selection, and dual-graph propagation with entropy regularization. Experiments demonstrate that MOSS achieves a superior accuracy--diversity trade-off over strong baselines.

\section*{Acknowledgments}
This work was partially supported by the National Natural Science Foundation of China under Project No. 62073272; the Guangdong Basic and Applied Basic Research Foundation under Grant No. 2024A1515011949 and 2026A1515011672; the Shenzhen Science and Technology Program under Grant No. GXWD20231130110308001, JCYJ20250604145617023, JCYJ20240813104843058, \allowbreak JCYJ202408131048\newline37050, and SYSPG20241211173609009; the Guangdong Province Higher Education Research and Reform Project under Grant No. YueJiaoGaoHan (2024) No. 9 (1227); and the Guangdong Province General Colleges and Universities Innovation Team Project under No. 2022KCXTD038.

\bibliographystyle{ACM-Reference-Format}
\bibliography{sample-base}


\begin{thebibliography}{43}


\ifx \showCODEN    \undefined \def \showCODEN     #1{\unskip}     \fi
\ifx \showISBNx    \undefined \def \showISBNx     #1{\unskip}     \fi
\ifx \showISBNxiii \undefined \def \showISBNxiii  #1{\unskip}     \fi
\ifx \showISSN     \undefined \def \showISSN      #1{\unskip}     \fi
\ifx \showLCCN     \undefined \def \showLCCN      #1{\unskip}     \fi
\ifx \shownote     \undefined \def \shownote      #1{#1}          \fi
\ifx \showarticletitle \undefined \def \showarticletitle #1{#1}   \fi
\ifx \showURL      \undefined \def \showURL       {\relax}        \fi
\providecommand\bibfield[2]{#2}
\providecommand\bibinfo[2]{#2}
\providecommand\natexlab[1]{#1}
\providecommand\showeprint[2][]{arXiv:#2}

\bibitem[Areeb et~al\mbox{.}(2023)]%
        {areeb2023filter}
\bibfield{author}{\bibinfo{person}{Qazi~Mohammad Areeb}, \bibinfo{person}{Mohammad Nadeem}, \bibinfo{person}{Shahab~Saquib Sohail}, \bibinfo{person}{Raza Imam}, \bibinfo{person}{Faiyaz Doctor}, \bibinfo{person}{Yassine Himeur}, \bibinfo{person}{Amir Hussain}, {and} \bibinfo{person}{Abbes Amira}.} \bibinfo{year}{2023}\natexlab{}.
\newblock \showarticletitle{Filter bubbles in recommender systems: Fact or fallacy—A systematic review}.
\newblock \bibinfo{journal}{\emph{Wiley Interdisciplinary Reviews: Data Mining and Knowledge Discovery}} \bibinfo{volume}{13}, \bibinfo{number}{6} (\bibinfo{year}{2023}), \bibinfo{pages}{e1512}.
\newblock


\bibitem[Bai et~al\mbox{.}(2025)]%
        {bai2025qwen}
\bibfield{author}{\bibinfo{person}{Shuai Bai}, \bibinfo{person}{Keqin Chen}, \bibinfo{person}{Xuejing Liu}, \bibinfo{person}{Jialin Wang}, \bibinfo{person}{Wenbin Ge}, \bibinfo{person}{Sibo Song}, \bibinfo{person}{Kai Dang}, \bibinfo{person}{Peng Wang}, \bibinfo{person}{Shijie Wang}, \bibinfo{person}{Jun Tang}, \bibinfo{person}{Humen Zhong}, \bibinfo{person}{Yuanzhi Zhu}, \bibinfo{person}{Mingkun Yang}, \bibinfo{person}{Zhaohai Li}, \bibinfo{person}{Jianqiang Wan}, \bibinfo{person}{Pengfei Wang}, \bibinfo{person}{Wei Ding}, \bibinfo{person}{Zheren Fu}, \bibinfo{person}{Yiheng Xu}, \bibinfo{person}{Jiabo Ye}, \bibinfo{person}{Xi Zhang}, \bibinfo{person}{Tianbao Xie}, \bibinfo{person}{Zesen Cheng}, \bibinfo{person}{Hang Zhang}, \bibinfo{person}{Zhibo Yang}, \bibinfo{person}{Haiyang Xu}, {and} \bibinfo{person}{Junyang Lin}.} \bibinfo{year}{2025}\natexlab{}.
\newblock \bibinfo{title}{Qwen2.5-VL Technical Report}.
\newblock
\showeprint[arxiv]{2502.13923}~[cs.CV]
\urldef\tempurl%
\url{https://arxiv.org/abs/2502.13923}
\showURL{%
\tempurl}


\bibitem[Chen et~al\mbox{.}(2024)]%
        {chen2024DCinCF}
\bibfield{author}{\bibinfo{person}{Huiyuan Chen}, \bibinfo{person}{Vivian Lai}, \bibinfo{person}{Hongye Jin}, \bibinfo{person}{Zhimeng Jiang}, \bibinfo{person}{Mahashweta Das}, {and} \bibinfo{person}{Xia Hu}.} \bibinfo{year}{2024}\natexlab{}.
\newblock \showarticletitle{Towards mitigating dimensional collapse of representations in collaborative filtering}. In \bibinfo{booktitle}{\emph{Proceedings of the 17th ACM international conference on web search and data mining}}. \bibinfo{pages}{106--115}.
\newblock


\bibitem[Girsang et~al\mbox{.}(2020)]%
        {girsang2020collaborative}
\bibfield{author}{\bibinfo{person}{AS Girsang}, \bibinfo{person}{B Al~Faruq}, \bibinfo{person}{HR Herlianto}, {and} \bibinfo{person}{S Simbolon}.} \bibinfo{year}{2020}\natexlab{}.
\newblock \showarticletitle{Collaborative recommendation system in users of anime films}. In \bibinfo{booktitle}{\emph{Journal of Physics: Conference Series}}, Vol.~\bibinfo{volume}{1566}. IOP Publishing, \bibinfo{pages}{012057}.
\newblock


\bibitem[Gong et~al\mbox{.}(2025)]%
        {gong2025divgcl}
\bibfield{author}{\bibinfo{person}{Wenwen Gong}, \bibinfo{person}{Yangliao Geng}, \bibinfo{person}{Dan Zhang}, \bibinfo{person}{Yifan Zhu}, \bibinfo{person}{Xiaolong Xu}, \bibinfo{person}{Haolong Xiang}, \bibinfo{person}{Amin Beheshti}, \bibinfo{person}{Xuyun Zhang}, {and} \bibinfo{person}{Lianyong Qi}.} \bibinfo{year}{2025}\natexlab{}.
\newblock \showarticletitle{DivGCL: a graph contrastive learning model for diverse recommendation}. In \bibinfo{booktitle}{\emph{Proceedings of the AAAI Conference on Artificial Intelligence}}, Vol.~\bibinfo{volume}{39}. \bibinfo{pages}{16853--16861}.
\newblock


\bibitem[Guo et~al\mbox{.}(2024)]%
        {guo2024lgmrec}
\bibfield{author}{\bibinfo{person}{Zhiqiang Guo}, \bibinfo{person}{Jianjun Li}, \bibinfo{person}{Guohui Li}, \bibinfo{person}{Chaoyang Wang}, \bibinfo{person}{Si Shi}, {and} \bibinfo{person}{Bin Ruan}.} \bibinfo{year}{2024}\natexlab{}.
\newblock \showarticletitle{Lgmrec: Local and global graph learning for multimodal recommendation}. In \bibinfo{booktitle}{\emph{Proceedings of the AAAI conference on artificial intelligence}}, Vol.~\bibinfo{volume}{38}. \bibinfo{pages}{8454--8462}.
\newblock


\bibitem[Hasan et~al\mbox{.}(2025)]%
        {hasan2025based}
\bibfield{author}{\bibinfo{person}{Emrul Hasan}, \bibinfo{person}{Mizanur Rahman}, \bibinfo{person}{Chen Ding}, \bibinfo{person}{Jimmy~Xiangji Huang}, {and} \bibinfo{person}{Shaina Raza}.} \bibinfo{year}{2025}\natexlab{}.
\newblock \showarticletitle{Based recommender systems: a survey of approaches, challenges and future perspectives}.
\newblock \bibinfo{journal}{\emph{Comput. Surveys}} \bibinfo{volume}{58}, \bibinfo{number}{1} (\bibinfo{year}{2025}), \bibinfo{pages}{1--41}.
\newblock


\bibitem[He and McAuley(2016)]%
        {he2016vbpr}
\bibfield{author}{\bibinfo{person}{Ruining He} {and} \bibinfo{person}{Julian McAuley}.} \bibinfo{year}{2016}\natexlab{}.
\newblock \showarticletitle{VBPR: visual Bayesian Personalized Ranking from implicit feedback}. In \bibinfo{booktitle}{\emph{Proceedings of the Thirtieth AAAI Conference on Artificial Intelligence}} (Phoenix, Arizona) \emph{(\bibinfo{series}{AAAI'16})}. \bibinfo{publisher}{AAAI Press}, \bibinfo{pages}{144–150}.
\newblock


\bibitem[He et~al\mbox{.}(2023)]%
        {He2023CGCL}
\bibfield{author}{\bibinfo{person}{Wei He}, \bibinfo{person}{Guohao Sun}, \bibinfo{person}{Jinhu Lu}, {and} \bibinfo{person}{Xiu~Susie Fang}.} \bibinfo{year}{2023}\natexlab{}.
\newblock \showarticletitle{Candidate-aware Graph Contrastive Learning for Recommendation}. In \bibinfo{booktitle}{\emph{Proceedings of the 46th International ACM SIGIR Conference on Research and Development in Information Retrieval}} (Taipei, Taiwan) \emph{(\bibinfo{series}{SIGIR '23})}. \bibinfo{publisher}{Association for Computing Machinery}, \bibinfo{address}{New York, NY, USA}, \bibinfo{pages}{1670–1679}.
\newblock
\showISBNx{9781450394086}
\href{https://doi.org/10.1145/3539618.3591647}{doi:\nolinkurl{10.1145/3539618.3591647}}


\bibitem[He et~al\mbox{.}(2020)]%
        {he2020lightgcn}
\bibfield{author}{\bibinfo{person}{Xiangnan He}, \bibinfo{person}{Kuan Deng}, \bibinfo{person}{Xiang Wang}, \bibinfo{person}{Yan Li}, \bibinfo{person}{Yongdong Zhang}, {and} \bibinfo{person}{Meng Wang}.} \bibinfo{year}{2020}\natexlab{}.
\newblock \showarticletitle{Lightgcn: Simplifying and powering graph convolution network for recommendation}. In \bibinfo{booktitle}{\emph{Proceedings of the 43rd International ACM SIGIR conference on research and development in Information Retrieval}}. \bibinfo{pages}{639--648}.
\newblock


\bibitem[Imran et~al\mbox{.}(2023)]%
        {imran2023refrs}
\bibfield{author}{\bibinfo{person}{Mubashir Imran}, \bibinfo{person}{Hongzhi Yin}, \bibinfo{person}{Tong Chen}, \bibinfo{person}{Quoc Viet~Hung Nguyen}, \bibinfo{person}{Alexander Zhou}, {and} \bibinfo{person}{Kai Zheng}.} \bibinfo{year}{2023}\natexlab{}.
\newblock \showarticletitle{ReFRS: Resource-efficient federated recommender system for dynamic and diversified user preferences}.
\newblock \bibinfo{journal}{\emph{ACM Transactions on Information Systems}} \bibinfo{volume}{41}, \bibinfo{number}{3} (\bibinfo{year}{2023}), \bibinfo{pages}{1--30}.
\newblock


\bibitem[Kang and McAuley(2018)]%
        {kang2018self}
\bibfield{author}{\bibinfo{person}{Wang-Cheng Kang} {and} \bibinfo{person}{Julian McAuley}.} \bibinfo{year}{2018}\natexlab{}.
\newblock \showarticletitle{Self-attentive sequential recommendation}. In \bibinfo{booktitle}{\emph{2018 IEEE international conference on data mining (ICDM)}}. IEEE, \bibinfo{pages}{197--206}.
\newblock


\bibitem[Liu et~al\mbox{.}(2024b)]%
        {liu2024drgame}
\bibfield{author}{\bibinfo{person}{Kangzhe Liu}, \bibinfo{person}{Jianghong Ma}, \bibinfo{person}{Shanshan Feng}, \bibinfo{person}{Haijun Zhang}, {and} \bibinfo{person}{Zhao Zhang}.} \bibinfo{year}{2024}\natexlab{b}.
\newblock \showarticletitle{DRGame: Diversified Recommendation for Multi-category Video Games with Balanced Implicit Preferences}. In \bibinfo{booktitle}{\emph{International Conference on Database Systems for Advanced Applications}}. Springer, \bibinfo{pages}{254--263}.
\newblock


\bibitem[Liu et~al\mbox{.}(2019)]%
        {Liu2019DC}
\bibfield{author}{\bibinfo{person}{Kanglin Liu}, \bibinfo{person}{Guoping Qiu}, \bibinfo{person}{Wenming Tang}, {and} \bibinfo{person}{Fei Zhou}.} \bibinfo{year}{2019}\natexlab{}.
\newblock \showarticletitle{Spectral Regularization for Combating Mode Collapse in GANs}. In \bibinfo{booktitle}{\emph{2019 IEEE/CVF International Conference on Computer Vision (ICCV)}}. \bibinfo{pages}{6381--6389}.
\newblock
\href{https://doi.org/10.1109/ICCV.2019.00648}{doi:\nolinkurl{10.1109/ICCV.2019.00648}}


\bibitem[Liu et~al\mbox{.}(2024a)]%
        {liu2024multimodal}
\bibfield{author}{\bibinfo{person}{Qidong Liu}, \bibinfo{person}{Jiaxi Hu}, \bibinfo{person}{Yutian Xiao}, \bibinfo{person}{Xiangyu Zhao}, \bibinfo{person}{Jingtong Gao}, \bibinfo{person}{Wanyu Wang}, \bibinfo{person}{Qing Li}, {and} \bibinfo{person}{Jiliang Tang}.} \bibinfo{year}{2024}\natexlab{a}.
\newblock \showarticletitle{Multimodal recommender systems: A survey}.
\newblock \bibinfo{journal}{\emph{Comput. Surveys}} \bibinfo{volume}{57}, \bibinfo{number}{2} (\bibinfo{year}{2024}), \bibinfo{pages}{1--17}.
\newblock


\bibitem[Lopez-Avila and Du(2025)]%
        {lopez2025survey}
\bibfield{author}{\bibinfo{person}{Alejo Lopez-Avila} {and} \bibinfo{person}{Jinhua Du}.} \bibinfo{year}{2025}\natexlab{}.
\newblock \showarticletitle{A survey on large language models in multimodal recommender systems}.
\newblock \bibinfo{journal}{\emph{arXiv preprint arXiv:2505.09777}} (\bibinfo{year}{2025}).
\newblock


\bibitem[Ma et~al\mbox{.}(2026)]%
        {ma2026SGMT}
\bibfield{author}{\bibinfo{person}{Jianghong Ma}, \bibinfo{person}{Changran He}, \bibinfo{person}{Dezhao Yang}, \bibinfo{person}{Tianjun Wei}, \bibinfo{person}{Haijun Zhang}, {and} \bibinfo{person}{Xiaofeng Zhang}.} \bibinfo{year}{2026}\natexlab{}.
\newblock \showarticletitle{SGMT: Social Generating with Multiview-Guided Tuning In Recommender Systems}.
\newblock \bibinfo{journal}{\emph{Proceedings of the AAAI Conference on Artificial Intelligence}} \bibinfo{volume}{40}, \bibinfo{number}{18} (\bibinfo{date}{Mar.} \bibinfo{year}{2026}), \bibinfo{pages}{15501--15509}.
\newblock


\bibitem[Ma et~al\mbox{.}(2021)]%
        {ma2021unified}
\bibfield{author}{\bibinfo{person}{Yao Ma}, \bibinfo{person}{Xiaorui Liu}, \bibinfo{person}{Tong Zhao}, \bibinfo{person}{Yozen Liu}, \bibinfo{person}{Jiliang Tang}, {and} \bibinfo{person}{Neil Shah}.} \bibinfo{year}{2021}\natexlab{}.
\newblock \showarticletitle{A unified view on graph neural networks as graph signal denoising}. In \bibinfo{booktitle}{\emph{Proceedings of the 30th ACM international conference on information \& knowledge management}}. \bibinfo{pages}{1202--1211}.
\newblock


\bibitem[Pathak et~al\mbox{.}(2017)]%
        {pathak2017generating}
\bibfield{author}{\bibinfo{person}{Apurva Pathak}, \bibinfo{person}{Kshitiz Gupta}, {and} \bibinfo{person}{Julian McAuley}.} \bibinfo{year}{2017}\natexlab{}.
\newblock \showarticletitle{Generating and personalizing bundle recommendations on steam}. In \bibinfo{booktitle}{\emph{Proceedings of the 40th international ACM SIGIR conference on research and development in information retrieval}}. \bibinfo{pages}{1073--1076}.
\newblock


\bibitem[Pomo et~al\mbox{.}(2025)]%
        {pomo2025recommender}
\bibfield{author}{\bibinfo{person}{Claudio Pomo}, \bibinfo{person}{Matteo Attimonelli}, \bibinfo{person}{Danilo Danese}, \bibinfo{person}{Fedelucio Narducci}, {and} \bibinfo{person}{Tommaso Di~Noia}.} \bibinfo{year}{2025}\natexlab{}.
\newblock \showarticletitle{Do recommender systems really leverage multimodal content? a comprehensive analysis on multimodal representations for recommendation}. In \bibinfo{booktitle}{\emph{Proceedings of the 34th ACM International Conference on Information and Knowledge Management}}. \bibinfo{pages}{2377--2387}.
\newblock


\bibitem[Putri and Faisal(2023)]%
        {putri2023analyzing}
\bibfield{author}{\bibinfo{person}{Helmy~Dianti Putri} {and} \bibinfo{person}{Muhammad Faisal}.} \bibinfo{year}{2023}\natexlab{}.
\newblock \showarticletitle{Analyzing the effectiveness of collaborative filtering and content-based filtering methods in anime recommendation systems}.
\newblock \bibinfo{journal}{\emph{Jurnal Komtika (Komputasi dan Informatika)}} \bibinfo{volume}{7}, \bibinfo{number}{2} (\bibinfo{year}{2023}), \bibinfo{pages}{124--133}.
\newblock


\bibitem[Raza et~al\mbox{.}(2026)]%
        {raza2026comprehensive}
\bibfield{author}{\bibinfo{person}{Shaina Raza}, \bibinfo{person}{Mizanur Rahman}, \bibinfo{person}{Safiullah Kamawal}, \bibinfo{person}{Armin Toroghi}, \bibinfo{person}{Ananya Raval}, \bibinfo{person}{Farshad Navah}, {and} \bibinfo{person}{Amirmohammad Kazemeini}.} \bibinfo{year}{2026}\natexlab{}.
\newblock \showarticletitle{A comprehensive review of recommender systems: Transitioning from theory to practice}.
\newblock \bibinfo{journal}{\emph{Computer Science Review}}  \bibinfo{volume}{59} (\bibinfo{year}{2026}), \bibinfo{pages}{100849}.
\newblock


\bibitem[Reimers and Gurevych(2019)]%
        {reimers2019sentencebert}
\bibfield{author}{\bibinfo{person}{Nils Reimers} {and} \bibinfo{person}{Iryna Gurevych}.} \bibinfo{year}{2019}\natexlab{}.
\newblock \showarticletitle{Sentence-bert: Sentence embeddings using siamese bert-networks}. In \bibinfo{booktitle}{\emph{Proceedings of the 2019 conference on empirical methods in natural language processing and the 9th international joint conference on natural language processing (EMNLP-IJCNLP)}}. \bibinfo{pages}{3982--3992}.
\newblock


\bibitem[Rendle et~al\mbox{.}(2009)]%
        {Rendle2009BPR}
\bibfield{author}{\bibinfo{person}{Steffen Rendle}, \bibinfo{person}{Christoph Freudenthaler}, \bibinfo{person}{Zeno Gantner}, {and} \bibinfo{person}{Lars Schmidt-Thieme}.} \bibinfo{year}{2009}\natexlab{}.
\newblock \showarticletitle{BPR: Bayesian personalized ranking from implicit feedback}. In \bibinfo{booktitle}{\emph{Proceedings of the Twenty-Fifth Conference on Uncertainty in Artificial Intelligence}} (Montreal, Quebec, Canada) \emph{(\bibinfo{series}{UAI '09})}. \bibinfo{publisher}{AUAI Press}, \bibinfo{address}{Arlington, Virginia, USA}, \bibinfo{pages}{452–461}.
\newblock
\showISBNx{9780974903958}


\bibitem[Roy and Vetterli(2007)]%
        {Roy2007effectiverank}
\bibfield{author}{\bibinfo{person}{Olivier Roy} {and} \bibinfo{person}{Martin Vetterli}.} \bibinfo{year}{2007}\natexlab{}.
\newblock \showarticletitle{The effective rank: A measure of effective dimensionality}. In \bibinfo{booktitle}{\emph{2007 15th European Signal Processing Conference}}. \bibinfo{pages}{606--610}.
\newblock


\bibitem[Wan and McAuley(2018)]%
        {wan2018item}
\bibfield{author}{\bibinfo{person}{Mengting Wan} {and} \bibinfo{person}{Julian McAuley}.} \bibinfo{year}{2018}\natexlab{}.
\newblock \showarticletitle{Item recommendation on monotonic behavior chains}. In \bibinfo{booktitle}{\emph{Proceedings of the 12th ACM conference on recommender systems}}. \bibinfo{pages}{86--94}.
\newblock


\bibitem[Wang and Isola(2020)]%
        {Wang2020UA}
\bibfield{author}{\bibinfo{person}{Tongzhou Wang} {and} \bibinfo{person}{Phillip Isola}.} \bibinfo{year}{2020}\natexlab{}.
\newblock \showarticletitle{Understanding contrastive representation learning through alignment and uniformity on the hypersphere}. In \bibinfo{booktitle}{\emph{Proceedings of the 37th International Conference on Machine Learning}} \emph{(\bibinfo{series}{ICML'20})}. \bibinfo{publisher}{JMLR.org}, Article \bibinfo{articleno}{921}, \bibinfo{numpages}{11}~pages.
\newblock


\bibitem[Wei et~al\mbox{.}(2020)]%
        {wei2020GRCN}
\bibfield{author}{\bibinfo{person}{Yinwei Wei}, \bibinfo{person}{Xiang Wang}, \bibinfo{person}{Liqiang Nie}, \bibinfo{person}{Xiangnan He}, {and} \bibinfo{person}{Tat-Seng Chua}.} \bibinfo{year}{2020}\natexlab{}.
\newblock \showarticletitle{Graph-refined convolutional network for multimedia recommendation with implicit feedback}. In \bibinfo{booktitle}{\emph{Proceedings of the 28th ACM international conference on multimedia}}. \bibinfo{pages}{3541--3549}.
\newblock


\bibitem[Xu et~al\mbox{.}(2026)]%
        {xu2026survey}
\bibfield{author}{\bibinfo{person}{Jinfeng Xu}, \bibinfo{person}{Zheyu Chen}, \bibinfo{person}{Shuo Yang}, \bibinfo{person}{Jinze Li}, \bibinfo{person}{Wei Wang}, \bibinfo{person}{Xiping Hu}, \bibinfo{person}{Steven Hoi}, {and} \bibinfo{person}{Edith Ngai}.} \bibinfo{year}{2026}\natexlab{}.
\newblock \showarticletitle{A survey on multimodal recommender systems: Recent advances and future directions}.
\newblock \bibinfo{journal}{\emph{IEEE Transactions on Multimedia}} (\bibinfo{year}{2026}).
\newblock


\bibitem[Yang et~al\mbox{.}(2023)]%
        {dgrec}
\bibfield{author}{\bibinfo{person}{Liangwei Yang}, \bibinfo{person}{Shengjie Wang}, \bibinfo{person}{Yunzhe Tao}, \bibinfo{person}{Jiankai Sun}, \bibinfo{person}{Xiaolong Liu}, \bibinfo{person}{Philip~S Yu}, {and} \bibinfo{person}{Taiqing Wang}.} \bibinfo{year}{2023}\natexlab{}.
\newblock \showarticletitle{Dgrec: Graph neural network for recommendation with diversified embedding generation}. In \bibinfo{booktitle}{\emph{Proceedings of the sixteenth ACM international conference on web search and data mining}}. \bibinfo{pages}{661--669}.
\newblock


\bibitem[Yu et~al\mbox{.}(2022)]%
        {liu2024simgcl}
\bibfield{author}{\bibinfo{person}{Junliang Yu}, \bibinfo{person}{Hongzhi Yin}, \bibinfo{person}{Xin Xia}, \bibinfo{person}{Tong Chen}, \bibinfo{person}{Lizhen Cui}, {and} \bibinfo{person}{Quoc Viet~Hung Nguyen}.} \bibinfo{year}{2022}\natexlab{}.
\newblock \showarticletitle{Are Graph Augmentations Necessary? Simple Graph Contrastive\&nbsp;Learning for Recommendation}. In \bibinfo{booktitle}{\emph{Proceedings of the 45th International ACM SIGIR Conference on Research and Development in Information Retrieval}} (Madrid, Spain) \emph{(\bibinfo{series}{SIGIR '22})}. \bibinfo{publisher}{Association for Computing Machinery}, \bibinfo{address}{New York, NY, USA}, \bibinfo{pages}{1294–1303}.
\newblock
\showISBNx{9781450387323}
\href{https://doi.org/10.1145/3477495.3531937}{doi:\nolinkurl{10.1145/3477495.3531937}}


\bibitem[Yu et~al\mbox{.}(2023)]%
        {yu2023multi}
\bibfield{author}{\bibinfo{person}{Penghang Yu}, \bibinfo{person}{Zhiyi Tan}, \bibinfo{person}{Guanming Lu}, {and} \bibinfo{person}{Bing-Kun Bao}.} \bibinfo{year}{2023}\natexlab{}.
\newblock \showarticletitle{Multi-view graph convolutional network for multimedia recommendation}. In \bibinfo{booktitle}{\emph{Proceedings of the 31st ACM international conference on multimedia}}. \bibinfo{pages}{6576--6585}.
\newblock


\bibitem[Yuan et~al\mbox{.}(2023)]%
        {yuan2023wheregonext}
\bibfield{author}{\bibinfo{person}{Zheng Yuan}, \bibinfo{person}{Fajie Yuan}, \bibinfo{person}{Yu Song}, \bibinfo{person}{Youhua Li}, \bibinfo{person}{Junchen Fu}, \bibinfo{person}{Fei Yang}, \bibinfo{person}{Yunzhu Pan}, {and} \bibinfo{person}{Yongxin Ni}.} \bibinfo{year}{2023}\natexlab{}.
\newblock \showarticletitle{Where to Go Next for Recommender Systems? ID- vs. Modality-based Recommender Models Revisited}. In \bibinfo{booktitle}{\emph{Proceedings of the 46th International ACM SIGIR Conference on Research and Development in Information Retrieval}} (Taipei, Taiwan) \emph{(\bibinfo{series}{SIGIR '23})}. \bibinfo{publisher}{Association for Computing Machinery}, \bibinfo{address}{New York, NY, USA}, \bibinfo{pages}{2639–2649}.
\newblock
\showISBNx{9781450394086}
\href{https://doi.org/10.1145/3539618.3591932}{doi:\nolinkurl{10.1145/3539618.3591932}}


\bibitem[Zhang et~al\mbox{.}(2025b)]%
        {zhang2025dual}
\bibfield{author}{\bibinfo{person}{Jingmao Zhang}, \bibinfo{person}{Zhiting Zhao}, \bibinfo{person}{Yunqi Lin}, \bibinfo{person}{Jianghong Ma}, \bibinfo{person}{Tianjun Wei}, \bibinfo{person}{Haijun Zhang}, {and} \bibinfo{person}{Xiaofeng Zhang}.} \bibinfo{year}{2025}\natexlab{b}.
\newblock \showarticletitle{Dual-Phase Playtime-guided Recommendation: Interest Intensity Exploration and Multimodal Random Walks}. In \bibinfo{booktitle}{\emph{Proceedings of the 33rd ACM International Conference on Multimedia}}. \bibinfo{pages}{6232--6241}.
\newblock


\bibitem[Zhang et~al\mbox{.}(2024)]%
        {zhang2024dual}
\bibfield{author}{\bibinfo{person}{Lingzi Zhang}, \bibinfo{person}{Xin Zhou}, \bibinfo{person}{Zhiwei Zeng}, {and} \bibinfo{person}{Zhiqi Shen}.} \bibinfo{year}{2024}\natexlab{}.
\newblock \showarticletitle{Dual-view whitening on pre-trained text embeddings for sequential recommendation}. In \bibinfo{booktitle}{\emph{Proceedings of the AAAI Conference on Artificial Intelligence}}, Vol.~\bibinfo{volume}{38}. \bibinfo{pages}{9332--9340}.
\newblock


\bibitem[Zhang et~al\mbox{.}(2025a)]%
        {zhang2025freelunch}
\bibfield{author}{\bibinfo{person}{Weizhi Zhang}, \bibinfo{person}{Liangwei Yang}, \bibinfo{person}{Wooseong Yang}, \bibinfo{person}{Henry~Peng Zou}, \bibinfo{person}{Yuqing Liu}, \bibinfo{person}{Ke Xu}, \bibinfo{person}{Sourav Medya}, {and} \bibinfo{person}{Philip~S Yu}.} \bibinfo{year}{2025}\natexlab{a}.
\newblock \showarticletitle{Llminit: A free lunch from large language models for selective initialization of recommendation}. In \bibinfo{booktitle}{\emph{Proceedings of the 2025 Conference on Empirical Methods in Natural Language Processing: Industry Track}}. \bibinfo{pages}{2016--2024}.
\newblock


\bibitem[Zhang et~al\mbox{.}(2023)]%
        {Zhang2023logdet}
\bibfield{author}{\bibinfo{person}{Yifei Zhang}, \bibinfo{person}{Hao Zhu}, \bibinfo{person}{yankai Chen}, \bibinfo{person}{Zixing Song}, \bibinfo{person}{Piotr Koniusz}, {and} \bibinfo{person}{Irwin King}.} \bibinfo{year}{2023}\natexlab{}.
\newblock \showarticletitle{Mitigating the Popularity Bias of Graph Collaborative Filtering: A Dimensional Collapse Perspective}. In \bibinfo{booktitle}{\emph{Advances in Neural Information Processing Systems}}, \bibfield{editor}{\bibinfo{person}{A.~Oh}, \bibinfo{person}{T.~Naumann}, \bibinfo{person}{A.~Globerson}, \bibinfo{person}{K.~Saenko}, \bibinfo{person}{M.~Hardt}, {and} \bibinfo{person}{S.~Levine}} (Eds.), Vol.~\bibinfo{volume}{36}. \bibinfo{publisher}{Curran Associates, Inc.}, \bibinfo{pages}{67533--67550}.
\newblock
\urldef\tempurl%
\url{https://proceedings.neurips.cc/paper_files/paper/2023/file/d5753be6f71fbfefaf47aa27ec41279c-Paper-Conference.pdf}
\showURL{%
\tempurl}


\bibitem[Zhang et~al\mbox{.}(2025c)]%
        {zhang2025DCinGCL}
\bibfield{author}{\bibinfo{person}{Yifei Zhang}, \bibinfo{person}{Hao Zhu}, \bibinfo{person}{Menglin Yang}, \bibinfo{person}{Jiahong Liu}, \bibinfo{person}{Rex Ying}, \bibinfo{person}{Irwin King}, {and} \bibinfo{person}{Piotr Koniusz}.} \bibinfo{year}{2025}\natexlab{c}.
\newblock \showarticletitle{Understanding and mitigating hyperbolic dimensional collapse in graph contrastive learning}. In \bibinfo{booktitle}{\emph{Proceedings of the 31st ACM SIGKDD Conference on Knowledge Discovery and Data Mining V. 1}}. \bibinfo{pages}{1984--1995}.
\newblock


\bibitem[Zhao et~al\mbox{.}(2025)]%
        {zhao2025fairness}
\bibfield{author}{\bibinfo{person}{Yuying Zhao}, \bibinfo{person}{Yu Wang}, \bibinfo{person}{Yunchao Liu}, \bibinfo{person}{Xueqi Cheng}, \bibinfo{person}{Charu~C Aggarwal}, {and} \bibinfo{person}{Tyler Derr}.} \bibinfo{year}{2025}\natexlab{}.
\newblock \showarticletitle{Fairness and diversity in recommender systems: a survey}.
\newblock \bibinfo{journal}{\emph{ACM Transactions on Intelligent Systems and Technology}} \bibinfo{volume}{16}, \bibinfo{number}{1} (\bibinfo{year}{2025}), \bibinfo{pages}{1--28}.
\newblock


\bibitem[Zheng et~al\mbox{.}(2021)]%
        {dgcn}
\bibfield{author}{\bibinfo{person}{Yu Zheng}, \bibinfo{person}{Chen Gao}, \bibinfo{person}{Liang Chen}, \bibinfo{person}{Depeng Jin}, {and} \bibinfo{person}{Yong Li}.} \bibinfo{year}{2021}\natexlab{}.
\newblock \showarticletitle{DGCN: Diversified recommendation with graph convolutional networks}. In \bibinfo{booktitle}{\emph{Proceedings of the web conference 2021}}. \bibinfo{pages}{401--412}.
\newblock


\bibitem[Zhou and Shen(2023)]%
        {zhou2023tale}
\bibfield{author}{\bibinfo{person}{Xin Zhou} {and} \bibinfo{person}{Zhiqi Shen}.} \bibinfo{year}{2023}\natexlab{}.
\newblock \showarticletitle{A tale of two graphs: Freezing and denoising graph structures for multimodal recommendation}. In \bibinfo{booktitle}{\emph{Proceedings of the 31st ACM international conference on multimedia}}. \bibinfo{pages}{935--943}.
\newblock


\bibitem[Zhou et~al\mbox{.}(2025)]%
        {zhou2025learning}
\bibfield{author}{\bibinfo{person}{Xin Zhou}, \bibinfo{person}{Xiaoxiong Zhang}, \bibinfo{person}{Dusit Niyato}, {and} \bibinfo{person}{Zhiqi Shen}.} \bibinfo{year}{2025}\natexlab{}.
\newblock \bibinfo{title}{Learning Item Representations Directly from Multimodal Features for Effective Recommendation}.
\newblock
\showeprint[arxiv]{2505.04960}~[cs.IR]
\urldef\tempurl%
\url{https://arxiv.org/abs/2505.04960}
\showURL{%
\tempurl}


\bibitem[Zhou et~al\mbox{.}(2023)]%
        {zhou2023TMFUN}
\bibfield{author}{\bibinfo{person}{Yan Zhou}, \bibinfo{person}{Jie Guo}, \bibinfo{person}{Hao Sun}, \bibinfo{person}{Bin Song}, {and} \bibinfo{person}{Fei~Richard Yu}.} \bibinfo{year}{2023}\natexlab{}.
\newblock \showarticletitle{Attention-guided Multi-step Fusion: A Hierarchical Fusion Network for Multimodal Recommendation}. In \bibinfo{booktitle}{\emph{Proceedings of the 46th International ACM SIGIR Conference on Research and Development in Information Retrieval}} (Taipei, Taiwan) \emph{(\bibinfo{series}{SIGIR '23})}. \bibinfo{publisher}{Association for Computing Machinery}, \bibinfo{address}{New York, NY, USA}, \bibinfo{pages}{1816–1820}.
\newblock
\showISBNx{9781450394086}
\href{https://doi.org/10.1145/3539618.3591950}{doi:\nolinkurl{10.1145/3539618.3591950}}


\end{thebibliography}


\end{document}